\documentclass[aps,pre,twocolumn,superscriptaddress,preprintnumbers,nofootinbib]{revtex4-2}
\usepackage{graphicx,epstopdf,amssymb,amsmath,verbatim,hyperref,mathdots,xcolor}
\pdfoutput=1

\newcommand{\kobe}[1]{{{\textbf{[KL:\ #1]}}}}
\newcommand{\wati}[1]{{{\textbf{[WT:\ #1]}}}}
\newcommand{\sarah}[1]{{{\textbf{[ZF:\ #1]}}}}
\newcommand{\mehran}[1]{{{\textbf{[MK:\ #1]}}}}

\newcommand{\clean}{
\renewcommand{\kobe}[1]{}
\renewcommand{\wati}[1]{}
\renewcommand{\sarah}[1]{}
\renewcommand{\mehran}[1]{}
}
\clean

\begin{document}

\preprint{MIT-CTP/5922}

\title{Minimal model of self-organized clusters with phase transitions in ecological communities}

\author{Shing Yan Li}
\email{sykobeli@mit.edu}

\affiliation{MIT Center for Theoretical Physics - a Leinweber Institute,  Cambridge, MA 02139, USA}

\author{Mehran Kardar}
\email{kardar@mit.edu}

\affiliation{Department of Physics, Massachusetts Institute of Technology, Cambridge, MA 02139, USA}

\author{Zhijie Feng}
\email{zjfeng@bu.edu}

\affiliation{Department of Physics, Boston University, Boston, MA 02215, USA}

\author{Washington Taylor}
\email{wati@mit.edu}

\affiliation{MIT Center for Theoretical Physics - a Leinweber Institute,  Cambridge, MA 02139, USA}

\date{\today}

\begin{abstract}
In complex ecological communities, species may self-organize into clusters or
clumps where highly similar species can coexist. The emergence of such species
clusters can be captured by the interplay between neutral and niche theories.
Based on the generalized Lotka-Volterra model of competition, we propose a
minimal model for ecological communities in which the steady states contain
self-organized clusters. In this model, species compete only with their neighbors in niche space
through a common interaction strength. Unlike many previous theories,
this model does not rely on random heterogeneity in interactions. 
Even in this minimal model where only the common interaction 
strength is varied, we find an exponentially large set of states that exhibit a rich variety of cluster patterns with different sizes and
combinations.
There are sharp phase transitions into the formation of clusters.
There are also multiple phase transitions between different sets of
possible cluster patterns, many of which accumulate near a small
number of critical points.
We analyze  
this phase structure using both numerical and analytical methods.
In addition, the special case with only nearest neighbor interactions is exactly
solvable using the method of transfer matrices from statistical
mechanics. We analyze 
the critical behavior of these
systems.
\end{abstract}

\maketitle


\section{Introduction}
\label{sec:Introduction}

\begin{figure*}
    \centering
    \includegraphics[width=1\textwidth]{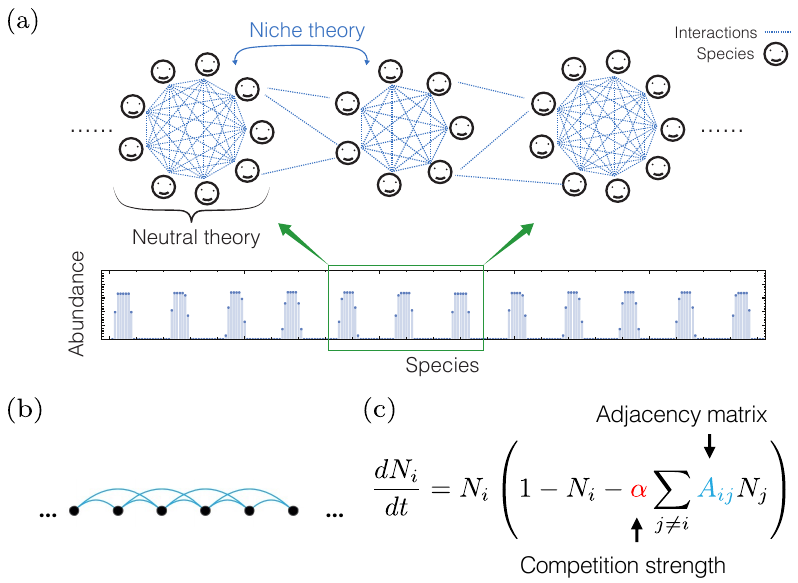}
    \caption{\textbf{Self-organized species clusters due to the interplay between neutral and niche theories, captured by a Lotka-Volterra model.} (a) The above illustrates cluster 
    patterns in species abundances along, for example, a niche axis, as well as
    the interactions among three of the clusters. On one hand, neutral theory predicts that
    multiple highly similar species can coexist within a cluster, with stronger or denser but balanced interactions. On the other 
    hand, niche theory predicts that different clusters can coexist only when
    they occupy sufficiently distinct niches, with weaker or sparser interactions
    in between. 
    (b) In our model of self-organized clusters, the interactions are 
    represented by the interaction network of 
    species, in which each edge
    has the same interaction strength $\alpha$.
    Each species is represented by a node in the 
    network, and only 
    interacts with its $2K$ ($=4$ in this example) nearest neighbors, as shown by the edges in the network. 
    (c)
    The abundances of these species $N_i$ follow the Lotka-Volterra dynamics (Eq. 
    \eqref{eq:GLV}) with competition strength $\alpha$. The adjacency matrix $A_{ij}$
    represents the interaction network.
    }
    \label{fig:clusterIntro}
\end{figure*}

In 
ecological systems, species are often more likely to coexist 
when they are either very similar or very different.
This tendency leads 
to the formation of clusters of species with similar traits or niches. Such
a
pattern is repeatedly observed in a wide range of ecological studies involving 
plants, animals, and plankton 
\cite{ClusterExperimentMammal,ClusterExperimentBird,ClusterExperimentLake,ClusterExperimentNewerPlankton}. 
Recent advances in metagenomic analysis further quantitatively confirm the 
presence of such clustering in microbiome communities at the strain level 
\cite{zhao2019adaptive,zheng2022high,jin2023culturing}. 
Explaining this phenomenon requires connecting two seemingly opposing ecological 
theories. On one hand, the competitive exclusion principle in niche theory 
suggests that in order to coexist, species must differ sufficiently in their 
niches, by occupying distinct habitats or utilizing different resources
\cite{Levins1962I, Levins1962II, MacArthurLimitingSimilarity, MacArthurPacking, TilmanResource}. It is evident that species tend to compete more strongly with 
those sharing similar trait and behavior \cite{lemos2024phylogeny}, reducing 
their likelihood to coexist. On the other hand, neutral theory 
\cite{HubbellNeutral, NeutralAgeTen, NeutralSM} argues that species with 
sufficiently similar ecological functions are effectively equivalent, allowing 
them to coexist in a manner analogous to populations of the same species. Such 
similarity often results from environmental filtering, where species sharing a 
habitat must adapt to the same environmental conditions. These two theories can 
both apply within the same community \cite{MehtaNicheNeutralTransition, GenralizedIsland}, and their interplay leads to the emergence of species clusters. See Fig. \ref{fig:clusterIntro} for an illustration of such species clusters. 

In theoretical
ecology, an early example of species clusters is the transient states in the 
one-dimensional niche axis model \cite{SchefferCluster, SchefferClusterExplained}. In the niche axis model \cite{WhittakerNicheAxis, WhittakerLevinNicheAxis}, species are ordered 
on a line according to their niches, and compete with one another following
Lotka-Volterra dynamics with a local, Gaussian-like interaction
kernel. While smooth interaction kernels can lead to clusters only in the
transient dynamics, it has been shown that non-smooth kernels can lead to 
clusters also in the steady states of the dynamics \cite{NonSmoothKernel, DeltaFunctionKernel}. We denote these clusters
in the steady states as \emph{self-organized clusters}.

Since then, the phenomenon of species
clusters has been identified in various ecological models \cite{ClusterRobust,ClusterRM,ClusterPredation}.
Nevertheless, a systematic theoretical framework for describing these clusters is still
lacking. In particular, the previous models were too complicated to be 
analytically tractable, thus the studies of these models were limited to numerical
simulations. Such limitation is an obstacle to classifying
possible self-organized cluster patterns, or identifying transitions between 
different sets of possible patterns, which may occur when characteristics of a 
community such as the
competition strength is changing. Tools
from statistical physics can help resolve some of these issues. Following the
terminology in statistical physics, we will refer to the transitions between sets
of patterns as phase transitions.

In this paper, we address some of the above issues by proposing a minimal model of
self-organized clusters based on Lotka-Volterra competitive dynamics with local
interactions. The model contains only three parameters: the number of species
$S$, the number of interacting neighbors $2K$, and the competition strength 
$\alpha$. In particular, even  
under the assumption
that all interactions between neighbors have 
the same strength, the steady states of such a simple model 
support an incredibly rich set of patterns that exhibit clusters of various sizes and combinations.
The model also has a rich phase diagram that arises by varying 
only $\alpha$. The phase 
diagram includes sharp phase transitions between states with and without clusters,
as well as multiple phase transitions between different sets of cluster patterns.
Interestingly, many of these phase transitions accumulate and become dense
near a small number of critical points. Among them, there is also a phase transition where 
long-range correlation between species abundances emerges.
Due to the simplicity of the model, many quantities
such as cluster sizes and the positions of the critical points are analytically
tractable. Moreover, the special case $K=1$, i.e., with nearest neighbor
interactions, is exactly solvable, enabling further 
analysis as a lattice model in statistical mechanics. 

The remainder of this paper is organized as follows. In Sec. \ref{sec:Model},
we introduce the generalized Lotka-Volterra model for self-organized clusters.
We also set up the notation and framework for describing steady states and 
cluster patterns in this model. After that, we describe in Sec. \ref{sec:Kequals1} the cluster patterns 
for the special case of nearest neighbor interactions, as an exactly solvable example.
Then, in Sec. \ref{sec:Analysis},
we describe the general phase diagram of this model and analyze each phase in detail.
We present results about the possible cluster patterns in different
phases and at the critical points. Many of these results are argued
using the fixed point conditions and linear stability. 
Finally, we conclude and discuss possible implications and extensions 
to our model in Sec. \ref{sec:Discussion}. 
The code for the simulations and figures in this paper can be found on \cite{githubCode}.

\section{Lotka-Volterra model}
\label{sec:Model}

We consider an ecological community assembled from a pool of $S$ competing species. To
study different patterns of species clusters in the long time scale, we focus on a
system
with a high number of
stable
steady states
containing distinct subsets of the species pool
(i.e., exhibiting multistability \cite{LotkaBook, MultistabilityBook}).
Each species
$i$ is characterized by its abundance $N_i$. The dynamics of the abundances
is described by a generalized Lotka-Volterra model:
\begin{equation}
\label{eq:GLV}
    \frac{dN_i}{dt}=N_i\left(1-N_i-\alpha\sum_{j\neq i}A_{ij}N_j\right)\,,
\end{equation}
where $\alpha>0$ is the competition
strength between species (or more precisely, the ratio between interspecific and
intraspecific competitions) \cite{LevinsEvolution,MayStabilityBook}, and $A_{ij}$ is the symmetric adjacency matrix of the
interaction network.
To focus on how
species interactions (instead of, e.g., intrinsic fitness differences) impact the cluster patterns, we
have set the carrying capacity to $1$ for
all species; we expect that heterogeneity in carrying capacities may significantly alter the exact 
abundance distribution across species, but  
 will not substantially affect some
more global features of the 
model such as the phase diagram.
The interaction network is characterized as a graph with interacting species
(nodes of the graph) connected by edges \cite{EcologicalNetwork}.
Species $i$ and $j$ can interact
with each other only if they are connected by an edge in the interaction 
network. The interaction strength $\alpha$ is the same for all interacting pairs
of species. As a minimal 
model of a network with local interactions, we consider species on a line (in niche space, for example) 
where each species is connected to 
its $2K$ nearest neighbors
($K$ on each side), see Fig. \ref{fig:clusterIntro}(b) for
an
example. We impose periodic boundary conditions on the network for simpler analysis.
Hence, the adjacency matrix is given by
\begin{equation}
\label{eq:adjacencyMatrix}
    A_{ij}=\begin{cases}
    1 & 1\leq d(i,j)\leq K\\
    0 & \mathrm{otherwise}
    \end{cases}\,,
\end{equation}
where $d(i,j)=\min\left\{|i-j|,S-|i-j|\right\}$ is the minimum distance between 
species $i$ and $j$.
Note that this interaction network is simply a regular ring network
with degree $2K$ \cite{albert2002statistical}.
The spatial description of the network, however, is more useful for connecting to 
more ecological contexts
such as niche axis or pattern formation. 
Below, we focus on the sparse limit $S \gg K \geq 1$, or the limit of large $S$ while keeping $K$ constant.

Since the interaction matrix $J_{ij}=\delta_{ij}+\alpha A_{ij}$ is symmetric, 
the model has
a Lyapunov function \cite{MacArthurPacking}
\begin{equation}
\label{eq:Lyapunov}
    E=-2\sum_i N_i + \sum_{ij}J_{ij}N_i N_j\,.
\end{equation}
The existence of
$E$ guarantees that the community reaches one of the stable fixed points given by local
minima of $E$ when restricted to nonnegative $N_i$.
We will be 
interested in the stable, uninvadable, and feasible fixed points of the model $N_i^*$, rather than the dynamics
reaching the fixed points. A fixed point is stable (in the direction of surviving species) if the submatrix of 
$J_{ij}$ for surviving species $i,j$ is positive definite.
A fixed point is uninvadable if the invasion fitnesses of extinct 
species $g_i(N_i^*=0)$ are negative, where the invasion fitness is defined as
\begin{equation}
    g_i=1-N_i-\alpha\sum_{j\neq i}A_{ij}N_j\,,
\end{equation}
that is the total growth rate in Eq. \eqref{eq:GLV}. A fixed point is feasible if $N_i^*\geq 0$ 
for all species. We note that at these fixed points, the value of $E$ is simply 
\begin{equation}
    E^*=-\sum_i N_i^*\,,
\end{equation}
i.e., the negative of the total biomass. 

In general, for a fixed $\alpha$ there are
many stable fixed points corresponding to different cluster
patterns.
To explore this set of fixed points, we use a combination of analytic and numerical 
simulation techniques; see Appendix \ref{sec:simulation} for the simulation method.

In most of the cases, when the community reaches a stable fixed point, the species organize
themselves into clusters of different sizes. Here, a cluster of size $n$ refers to a set
of $n$ surviving species at consecutive positions on the ring lattice, followed by a 
number of extinct species in the neighborhoods to the left and right. We call a set of $d$ extinct
species between two clusters a gap of size $d$. 


As a remark, our model is similar to the niche axis model with an exactly box-shaped
interaction kernel. In such a niche axis model, however, 
the intraspecific and interspecific
competitions are equally strong, corresponding to $\alpha=1$ in our model, so 
there are no clusters in the steady states. The difference between the intraspecific
and interspecific competitions can only be introduced with a delta function kernel 
\cite{DeltaFunctionKernel}.
It was argued in \cite{NonSmoothKernel} that this kind of discontinuity is unrealistic for a continuous family of species.
On the other hand, here we focus on discrete sets of species, thus we can assume finite differences between neighboring species and allow $\alpha$ to 
vary, which enables the rich structure of self-organized clusters and phase
transitions.

Since the model has a Lyapunov function $E$, our model also has a useful statistical 
mechanics variant, namely the canonical ensemble for the set of all 
stable fixed points based on the Lyapunov function $E^*$. 
In this ensemble, a fixed point with abundances $N_i^*$ occurs with probability 
following a Boltzmann distribution
\begin{equation}
\label{eq:canonicalEnsembleProb}
    p(N_i^*)\propto e^{-\beta E^*}=e^{\beta\sum_i N_i^*}\,,
\end{equation}
for some inverse temperature $\beta$. Such an ensemble arises when there is small
demographic noise in the dynamics \cite{BuninMarginalStability,DemographicNoise}, or when we consider a large set of random initial
conditions \cite{TaylorMultistability}. See Appendix \ref{sec:motivation} for a detailed discussion.

\section{Nearest neighbor interactions}
\label{sec:Kequals1}

Before discussing general results about the model in Sec.~\ref{sec:Model}, let us
focus on  
the particularly simple case of $K=1$, i.e., when the species only interact
with their two nearest neighbors. As we will see, the model with $K=1$ is exactly
solvable and we can explicitly write down all possible stable fixed points at any 
$\alpha$. Its statistical mechanics variant as in 
Eq. \eqref{eq:canonicalEnsembleProb} is also exactly solvable using the method of 
transfer matrices. These simplifications allow us to gain better intuition about the 
possible cluster patterns of the model and the corresponding phase diagram. For 
this
purpose, here we only summarize the main results by describing the set of cluster patterns, i.e., the stable fixed points for different $\alpha$, and give the full derivations in Appendix \ref{sec:moreKequals1}.

First, at $\alpha>1$, the interspecific competition is stronger than 
the intraspecific
competition, resulting in a competitive exclusion regime 
\cite{MacArthurLimitingSimilarity,MacArthurPacking}. Each surviving 
species is isolated from each other with no interacting neighbor, thus there is no species 
cluster. We further find that two adjacent
surviving species are separated by a gap with size either $d=1$ or $d=2$. Similar results 
for general $K$ will be proven in Sec.~\ref{sec:Analysis}.

Next, at $1/2<\alpha<1$, some of the surviving species become closer and interacting,
hence forming clusters, while the others remain isolated. The transition
point at $\alpha=1$ separates the regimes with and without clusters. We find that all
the gaps now have size $d=1$, and all the clusters with more than one surviving
species have the same size $n$ depending on $\alpha$. 
In Appendix \ref{subsec:uninvadable}, by analyzing the conditions on stability and
uninvadability, we show that $n=2l$ must be even and
$l$ is an integer satisfying the inequality
\begin{equation}
\label{eq:alphaInequality}
    \frac{1}{2\cos\frac{\pi}{2(l+1)+1}}\leq\alpha<\frac{1}{2\cos\frac{\pi}{2l+1}}\,.
\end{equation}
In particular, we see that as $\alpha$ decreases, there are multiple phase 
transitions where $l$ jumps up
by $1$. For example, the first transition at $\alpha=(\sqrt 5-1)/2\simeq 0.618$ goes from $l=1$ to $l=2$ as $\alpha$ decreases.

From Eq. \eqref{eq:alphaInequality}, we see that all these 
transition
points are above 
$\alpha=1/2$. In fact, in the large $S$ limit, many phase transitions 
accumulate near $\alpha=1/2$ as $l$ keeps growing. The accumulation continues until
we reach $\alpha\leq 1/2$, where $n$ reaches community size and all species coexist
uniformly.

Since a cluster with size $n=2l$ relates the
species abundances at its two ends, $l$ is analogous to the correlation
length $\xi$ in statistical mechanics, which indeed reaches system size at the critical
point of a phase transition. Accordingly, at $\alpha=1/2$,
long-range correlation 
between species abundances emerges, which is simply uniform coexistence for $K=1$
but, as we will see in Sec~\ref{subsec:alpha1/2}, is more complex in general for 
$K>1$. This analogy to correlation length can be made
explicit by considering the canonical ensemble as in 
Eq. \eqref{eq:canonicalEnsembleProb}, which is also exactly solvable using transfer matrices as done in 
Appendix \ref{sec:moreKequals1}.
Notably, we find that despite the presence of an arbitrary inverse temperature $\beta$, the 
resulting correlation length near $\alpha=1/2$ is independent of $\beta$ and instead
goes as
\begin{equation}
    \xi\sim l(\log l)^2\,,
\end{equation}
which is indeed close to $\xi\sim l$.

In conclusion, by analyzing the model with $K=1$, we have found multiple phase
transitions between the formation of clusters with various sizes depending on 
$\alpha<1$, and a critical 
point at $\alpha=1/2$ where phase transitions accumulate.

\section{The general phase diagram}
\label{sec:Analysis}

\begin{figure*}
    \centering
    \includegraphics[width=1\textwidth]{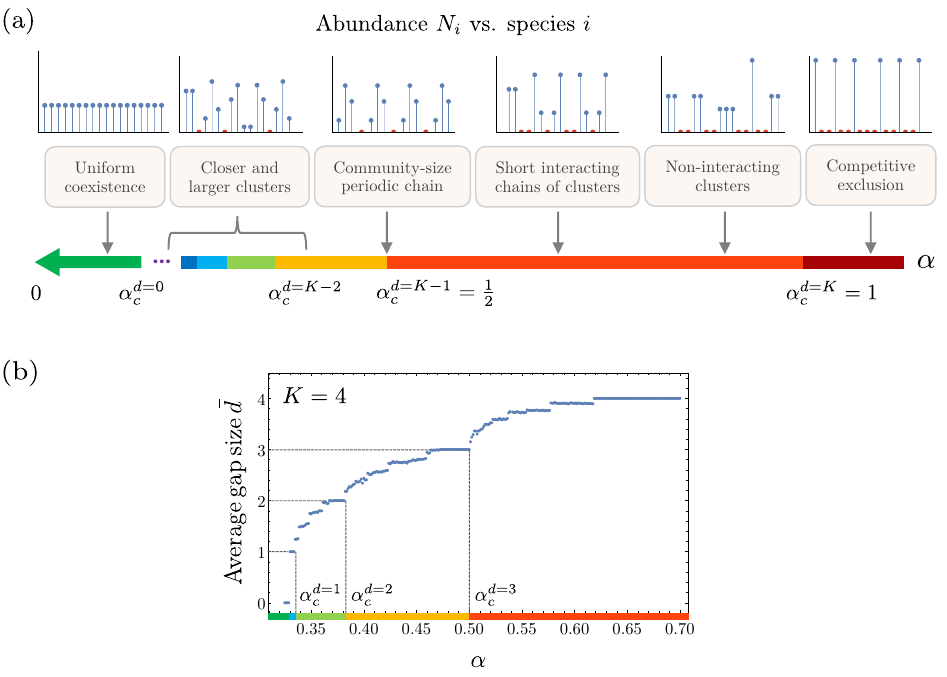}
    \caption{\textbf{General phase diagram of the Lotka-Volterra model with self-organized clusters.} 
      (a)
    The phase diagram of our model (Sec. \ref{sec:Analysis}) at different 
    competition strength $\alpha$. There is a critical point at $\alpha=1$ below
    which clusters start to emerge. There are also
    $K$ critical points where large number of phase transitions accumulate. These critical points are at $\alpha=\alpha_c^{d=0},\cdots,\alpha_c^{d=K-1}$, where $d$ denotes the number of extinct species 
    (referred to as gaps) between adjacent clusters in the
    interaction network at the corresponding critical point. The positions of the critical points are given by Eq. \eqref{eq:criticalPoints}. The plots of 
    species abundances show some of the possible cluster patterns in different
    phases, using $K=2$ as examples.
    The blue and red dots represent surviving and extinct species
    respectively.
    (b) The phase transitions and their accumulation at critical points are demonstrated by the average size of gaps as $\alpha$ 
    varies. Each accumulation starts from higher $\alpha$
    with a plateau at an integer average gap size, 
    i.e., all the
    gaps have the same size. As $\alpha$ decreases, chains of interacting 
    clusters emerge and cause discontinuous jumps in the average gap size.
    The jumps become denser until $\alpha$ reaches a critical point from the
    right. The gap sizes are obtained for $S=200,K=4$,
    and are sampled using random initial conditions. Each
    initial abundance is uniformly sampled between $0$ and $1$.
    }
    \label{fig:phasediagram}
\end{figure*}

We now turn to the general case with $K>1$.
As in $K=1$, when we vary $\alpha$, there are phase transitions between different types of cluster 
patterns. Many of the phase transitions are similar to those in $K=1$, but the full
phase diagram is more complex in general.

While there are many phase transitions, it will be useful to focus 
on the critical points where the 
maximum possible gap size between adjacent clusters $d_\mathrm{max}$ jumps between
discrete values.
In other words, we can treat $d_\mathrm{max}$ as an order parameter. More 
precisely, $d_\mathrm{max}$ jumps between $d$ (where $0\leq d\leq K-1$) and $d+1$ at
a critical point $\alpha=\alpha_c^d$. As we will see, there are dense
sets of phase transitions accumulating near these critical points, similar to the point $\alpha=1/2$ at $K=1$.
Similar phase structure for sparse and tree-like interaction networks has
been found in \cite{BuninSparse,TopologicalGlass}. 
We will use both analytical 
arguments and numerical simulations to study the 
properties of these phases. 

It is useful to first have an overview on the phase diagram as in Fig. 
\ref{fig:phasediagram}(a).
We start with $\alpha>1$ and gradually decrease $\alpha$. At first, there are 
not any clusters and each surviving
species is separated from the others with gaps of sizes $d\geq K$, hence is not 
interacting with any other species. After we reach $\alpha<1$, species start to form 
clusters with sizes from $1$ to $K+1$, but the clusters are separated with gap size $d=K$ and
still not interacting with each other. At $\alpha<(\sqrt 5 - 1)/2\simeq 0.618$ (which is independent of $K$),
some
clusters become closer to each other with $d=K-1$ instead of $d=K$. 
(For $K=1$, the gaps disappear and the clusters merge into larger ones with size $n=2l$.)
As a result,
there are small sets of clusters in which each cluster interacts with its neighboring
clusters. We refer to such configurations as chains of interacting
clusters. These chains grow longer as we decrease $\alpha$, 
until
$\alpha=1/2$ where the 
chains grow to community size. All clusters are now separated with $d=K-1$ and interacting with neighboring clusters. 
Therefore, $\alpha=1/2$ is again a critical point where phase transitions accumulate and long-range correlation of cluster
patterns emerges. For $K>1$, however, species do not uniformly coexist and there
are still clusters when $\alpha<1/2$.
Above $\alpha = 1/2$, the number of stable fixed points is exponential
in system size $S$, while for $\alpha \leq 1/2$, the number of stable
fixed points scales only polynomially in $S$.
(A more detailed quantitative description of the number of stable
fixed points and scaling rules is given in
Appendix~\ref{sec:scaling}.)

In contrast to $K=1$, there is an additional
cascade of phase transitions in the interacting regime at $\alpha<1/2$, although
there is always long-range correlation between clusters.
We start with a set of
interacting clusters separated by $d=k<K$ at a critical point. As we decrease $\alpha$, chains of clusters
with $d=k-1$ start growing, until another critical point where these chains grow
to community size, giving rise to a new phase with $d=k-1$.
This process 
repeats for different $d$ until we reach $d=0$, that is when all the species coexist with each other. In this regime, the community loses multistability and reaches
a unique fixed point. The value of $\alpha$ in this 
coexistence regime is of order $1/K$, and the model can be 
described by mean-field theory \cite{MayStability,BuninGLV,BuninMarginalStability}. Below we will analyze each of
these phases in detail.

\subsection{Onset of cluster formation at $\alpha=1$}
\label{subsec:Onset}

We first study the transition at $\alpha=1$, below which the species start to form
clusters; see the rightmost two plots in Fig. \ref{fig:phasediagram}(a). At $\alpha>1$, the interspecific competition is stronger than 
the intraspecific
competition, resulting in a competitive exclusion regime \cite{MacArthurLimitingSimilarity,MacArthurPacking}. Consider a single
surviving species with all
the $2K$ neighboring species interacting with it being extinct. The fixed point 
condition gives $N_i^*=1$ for the surviving species. Then we see that all the
neighboring species have invasion fitness $1-\alpha<0$ and cannot invade. 
On the other hand, any two interacting species have interaction matrix
\begin{equation}
    J_{ij}=\left(\begin{matrix}1&\alpha\\\alpha&1\end{matrix}\right)\,,
\end{equation}
which has a negative eigenvalue when $\alpha > 1$. Hence by the lemma in Appendix \ref{sec:simpleClusters}, all the surviving species must be non-interacting and separated by 
$d\geq K$.
The abundance is $N_i^*=1$ for all 
surviving species. The extinct species cannot invade as long as there is a surviving 
species in the neighborhood. On the other hand, two adjacent surviving species cannot be
separated 
by $d>2K$, otherwise the species at the middle of the gap does not interact with any 
surviving species, hence is invadable. Since a single surviving species is always a
stable configuration, we conclude that any configuration satisfying 
\begin{equation}
\label{eq:conditionAbove1}
    n=1\,,\quad K\leq d\leq 2K\,,
\end{equation}
is a possible final state of the community. See also \cite{TaylorMultistability} for
similar results in the niche axis model. From this condition, we see that the
total number of stable fixed points is exponential in $S$
(see Appendix~\ref{sec:scaling} for a more detailed computation).

In contrast, the community starts to form clusters when $\alpha<1$.
Consider a cluster with size $1\leq n\leq K+1$. When all the other species 
interacting with the cluster are extinct, the interaction submatrix is
\begin{equation}
    J_{ij}=\left(\begin{array}{cccc}
1 & \alpha & \cdots & \alpha\\
\alpha & 1 & \ddots & \vdots\\
\vdots & \ddots & \ddots & \alpha\\
\alpha & \cdots & \alpha & 1
\end{array}\right)\,,
\end{equation}
which has eigenvalues $1+(n-1)\alpha > 0$ and $1-\alpha > 0$, hence the
interaction subnetwork is stable when $\alpha<1$. 
Moreover, from the fixed point conditions,
all species in the cluster coexist with $N_i^*=1/(1+(n-1)\alpha)$, so the 
configuration is feasible. Therefore, the community can form clusters with size
$1\leq n\leq K+1$ at the stable fixed points.

The onset of cluster
formation can be intuitively understood as follows. When a cluster has size 
$1\leq n\leq K+1$, it has a fully-connected interaction subnetwork, and each 
species in the cluster competes
with other surviving species in exactly the same way \cite{CommunityClique}. 
The interspecific competition is perfectly balanced when the species
within a cluster have uniform abundance. Such a configuration is 
stabilized by the intraspecific competition if $\alpha <1$, i.e., when
the intraspecific
competition is stronger than the interspecific competition. 
This mechanism thus allows stable coexistence within a cluster.
On the other hand, if a cluster has size exceeding $K+1$, the interaction subnetwork is 
no longer fully-connected and the competition is no longer uniform. When $\alpha$ is 
sufficiently large but still smaller than $1$, the competition will always
drive some species to extinction, 
splitting the cluster into multiple smaller clusters. The same extinction occurs for sufficiently large 
$\alpha$ when we
bring two clusters to interact with each other at some gap size $d<K$;
see Sec. \ref{subsec:Chains}
for a more detailed description of
when interacting clusters start to coexist. In Appendix \ref{subsec:single}, we rigorously prove 
that for $1/2<\alpha<1$, the only possible sizes of clusters are indeed $1\leq n \leq K+1$.
Nevertheless, this bound can be violated for smaller $\alpha$.

We now turn to the gap sizes. We notice that any extinct species must interact with at least
two clusters. Suppose an extinct species only interacts with a single cluster. For $\alpha<1$, 
the invasion fitness of the extinct species is
\begin{subequations}
\begin{align}
    g_i&\geq 1-\alpha\sum_{j\in\mathrm{cluster}}N_j^*\\
    &>1-N_k^*-\alpha\sum_{\substack{j\neq k\\j\in\mathrm{cluster}}}N_j^*=g_k=0\,,
\end{align}
\end{subequations}
for any species $k$ in the cluster. Therefore, the 
extinct species can always invade if it only interacts with a single cluster. To avoid
such invasion, the gap size must be $d=K$ if the two clusters are not interacting with
each other. We conclude that the possible final states of the community for large 
$\alpha<1$ are non-interacting clusters 
with
\begin{equation}
\label{eq:conditionBelow1}
    1\leq n\leq K+1\,,\quad d=K\,,
\end{equation}
subject to uninvadability constraints, e.g., there cannot be two
adjacent clusters with size $n=K+1$.
Similar to the case of $\alpha>1$, the total number of stable fixed points is also 
exponential in $S$ for large $\alpha <1$. As we will see in Sec. \ref{subsec:alpha1/2},
this exponential behavior remains true for $\alpha>1/2$.


\subsection{Chains of interacting clusters at $\alpha>1/2$}
\label{subsec:Chains}

\begin{figure}
    \centering
    \includegraphics[width=1\columnwidth]{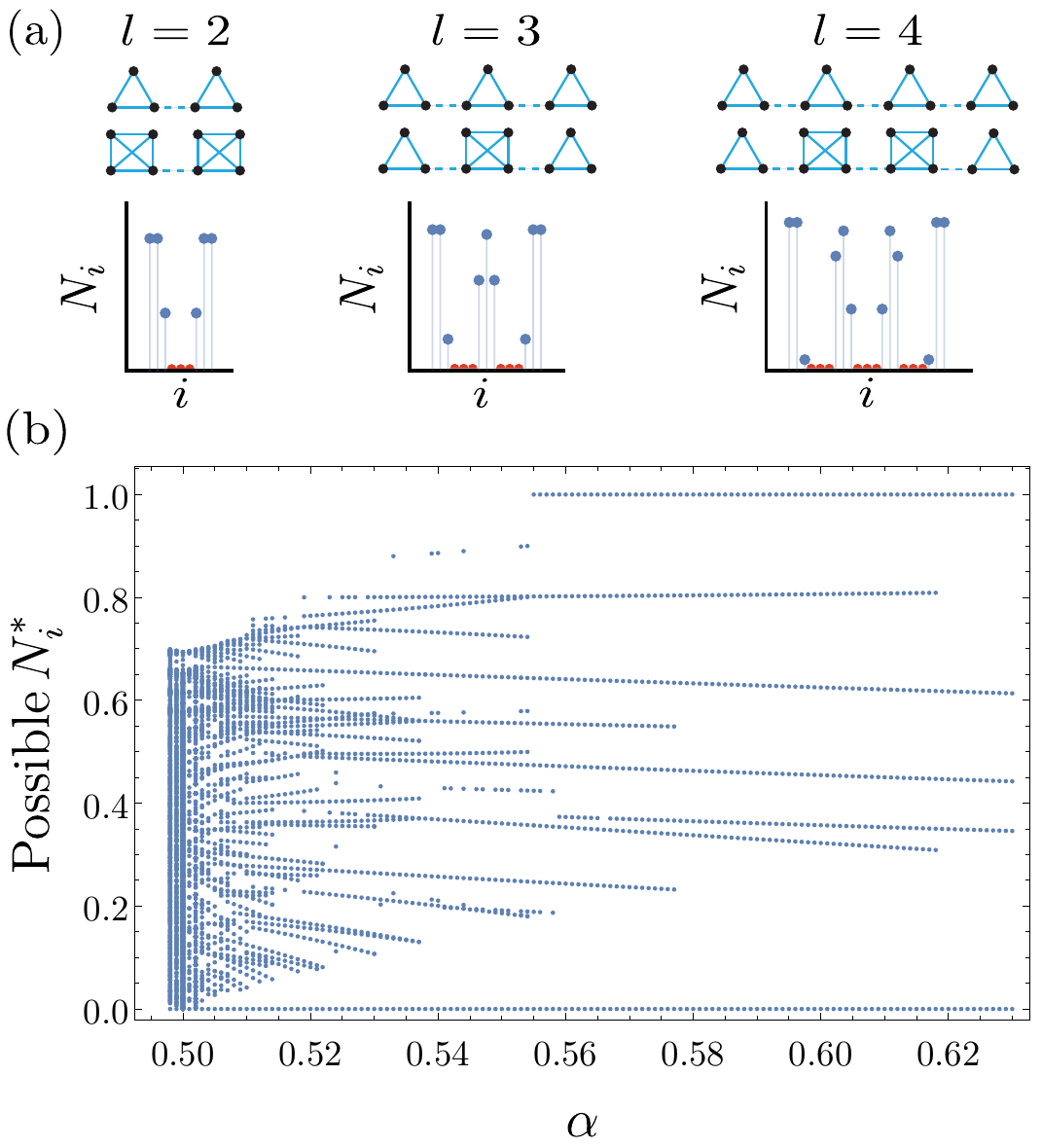}
    \caption{\textbf{Some possible chains of interacting clusters in the steady 
    states of the community when 
    $\alpha>1/2$.} (a) Some examples of chains of interacting clusters with
    lengths $2\leq l \leq 4$ when $K=4$. For each $l$, there are two examples
    of interaction networks containing $l$ clusters
    connected by single edges (dashed lines). There is also an example of species abundances of such chains for each $l$. (b) The bifurcation diagram for
    $K=4$ showing the possible values of abundances $N_i^*$ across different 
    stable fixed points at different $\alpha$. New lines emerge when $\alpha$ 
    decreases, representing phase transitions where new chain patterns become stable. The lines become denser 
    until
    the critical point at $\alpha=1/2$.
    }
    \label{fig:chains}
\end{figure}

As we continue to decrease $\alpha$ from $\alpha = 1$ to $\alpha = 1/2$,
some clusters become closer to each other and start
to interact while still coexisting. The minimal interaction between two clusters 
is achieved through a gap size of $d=K-1$, such that the two clusters are connected only
by a single edge in the interaction network. Indeed, for $\alpha>1/2$,
the possible  
patterns of interacting clusters are $l$ consecutive clusters of different sizes separated
by $d=K-1$; see Fig. \ref{fig:chains}(a) for examples. We refer to such patterns as chains of interacting 
clusters with length $l$. Since there is no interaction between adjacent chains,
the total number of stable fixed points is still exponential in $S$. 
Note that the species abundances in such chains are no longer
uniform within a cluster, but still have reflectional symmetries.

Fig. \ref{fig:chains}(b) shows the possible values of the abundances $N_i^*$ at different $\alpha$, obtained
through simulations with random sets of initial conditions. As we
decrease $\alpha$, phase transitions occur as new lines keep emerging at discrete values of $N_i^*$, signaling that
new chain patterns become stable and start appearing in the final state of the community.
The first chain pattern that becomes stable is a pair of clusters with size $n=2$, which 
becomes stable at $\alpha=(\sqrt 5 - 1)/2\simeq 0.618$. As a rigorous example, in Appendix \ref{subsec:pair} we show
that the possible patterns for $l=2$ are pairs of clusters with size $2\leq n\leq K$ 
separated by $d=K-1$, 
which become stable at
\begin{equation}
    \alpha_{n,l=2}=\frac{\sqrt{n+3}+\sqrt{n-1}}{\sqrt{n+3}+3\sqrt{n-1}}>\frac{1}{2}\,.
\end{equation}
Note that $\alpha_{n,l=2}$ decreases with $n$.

The possible chain length $l$ becomes larger for lower $\alpha$, since 
intuitively larger $l$ implies  a larger extent of coexistence, which requires a
smaller amount of competition. In fact, the critical 
$\alpha$ for a chain of length $l$ to become stable must satisfy the bound \cite{BuninSparse}
\begin{equation}
\label{eq:chainBound}
    \alpha_l\leq \frac{1}{2\cos\frac{\pi}{2l+1}}\,,
\end{equation}
which we prove in Appendix \ref{subsec:chainProof}.
Note that the bound becomes very close to $1/2$
for large $l$, so $\alpha_l$ becomes more dense as $l$ grows. As a result, a large number of new lines emerge in the plot of possible $N_i^*$ (Fig. \ref{fig:chains}(b)) when $\alpha$ is close to $1/2$. 

\subsection{The critical point at $\alpha=1/2$}
\label{subsec:alpha1/2}

The above analysis suggests that there is an accumulation of phase transitions at $\alpha=1/2$. We see that the
chains of interacting clusters grow to community size at $\alpha=1/2$, hence all clusters 
are separated by $d=K-1$ and interacting with neighboring clusters. As discussed in 
Sec.~\ref{sec:Kequals1}, the chain length is analogous to the correlation length in statistical mechanics.
We can also approximate the 
corresponding critical exponent: assuming that the bound in Eq.~\eqref{eq:chainBound}
is close to saturation for large $l$, we can then expand in $1/l$ and obtain
\begin{equation}
  \xi \sim l_\mathrm{max}\sim (\alpha-\alpha_c^{d=K-1})^{-1/2}\,,
  \label{eq:l-a}
\end{equation}
where $\alpha_c^{d=K-1}=1/2$ is the critical point.

The form of Eq.~\eqref{eq:l-a} suggests that the slope of the average
gap size as a function of $\alpha$ should diverge as the critical
point is approached from above.
If we assume that the average chain length
$\bar{l}(\alpha)$
is on the order of
$l_\mathrm{max}$ as $\alpha$ decreases, then the average gap size
$\bar{d}\sim K -1+ O(1/\bar{l}(\alpha))$
satisfies $d(\bar{d}(\alpha))/d \alpha \rightarrow \infty$ as $\alpha
\rightarrow \alpha_c^{d=K-1}$.
Additional numerical evidence for this conclusion regarding the sharp
nature of the phase transition appears in Fig.~\ref{fig:phasediagram}(b).
Note, however, that it is difficult to describe the behavior of
$\bar l$ or $\bar d$ more quantitatively, since they involve very complex rules of
possible chains of interacting clusters with different sizes $n\leq l_\mathrm{max}$, as well as the biased
sampling in this diverse set of cluster patterns due to random initial conditions (see also Appendix \ref{subsec:randomInitialConditions}).

\begin{figure*}
    \centering
    \includegraphics[width=1\linewidth]{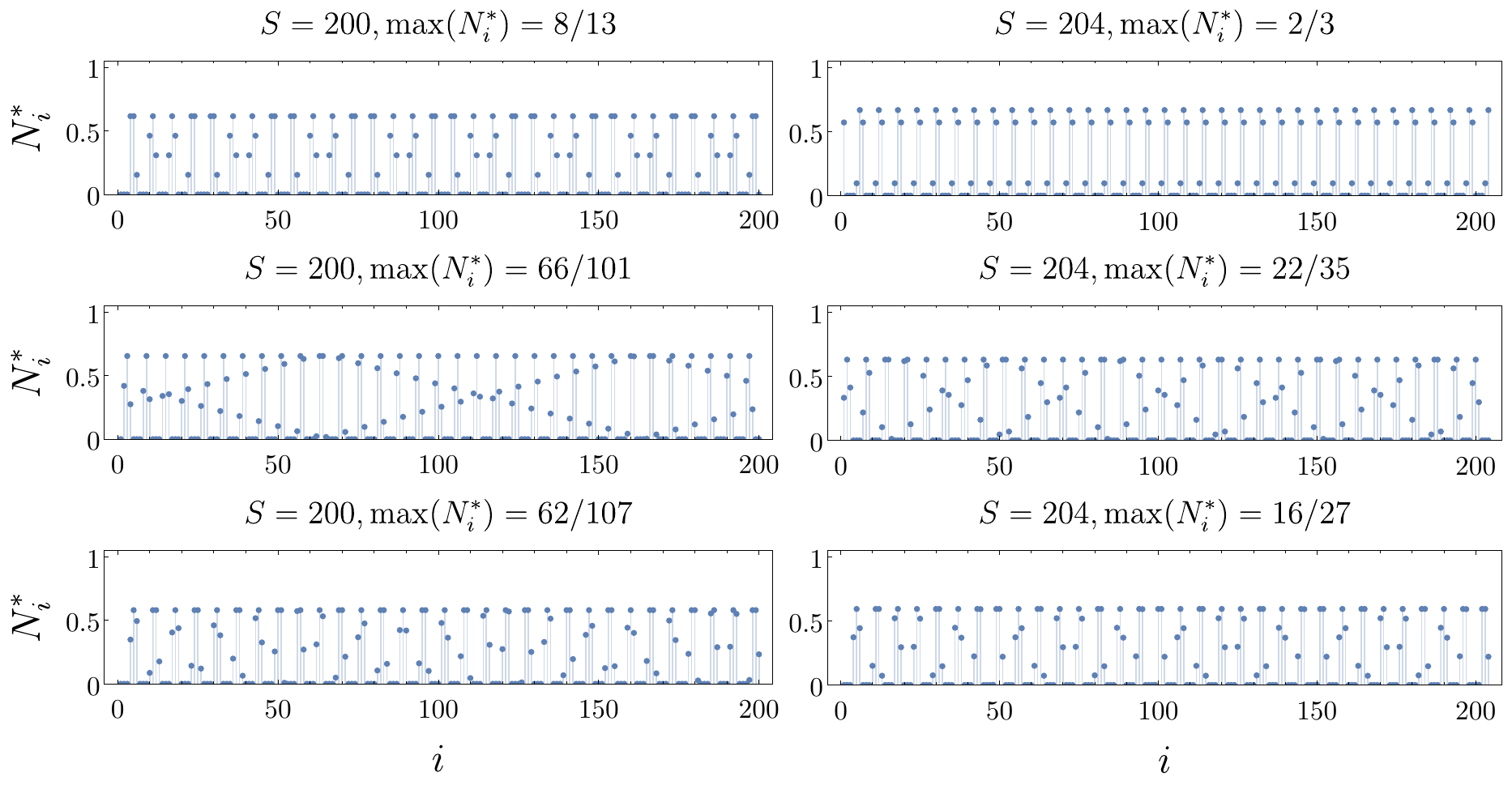}
    \caption{\textbf{Long-range correlation in cluster patterns at the critical point $\alpha=1/2$.} The examples of cluster patterns are with $K=4$ and $S=200,204$. 
    All clusters in each system 
    have the same maximum abundance $\max(N_i^*)$.
    For $\max(N_i^*)=66/101,62/107$, the patterns are only approximately periodic with periods shorter than $S$. All the other patterns are exactly periodic with periods shorter than $S$.
    }
    \label{fig:criticalPointExamples}
\end{figure*}

The cluster patterns at exactly $\alpha=1/2$ 
become
periodic and can be obtained
analytically. In Appendix \ref{sec:alpha1/2analysis}, we show that all clusters in
the community with size $n\geq 3$ share the same interior and maximum abundance
$z=\max(N_i^*)$. Interestingly, the choice of $z$ is sufficient to determine the 
sizes of all clusters and, hence, the period of the cluster pattern. To satisfy
the periodic boundary condition at finite $S$, $z$ must be a rational number in the
form of
\begin{equation}
    z=\frac{2p}{q}\,,
\end{equation}
where $p,q$ are integers satisfying
\begin{equation}
\label{eq:zCondition}
    (K-1)p+q=S\,.
\end{equation}
The period of the pattern is $S/\mathrm{gcd}(p,q)$. In addition, stability and uninvadability requires that
\begin{equation}
\label{eq:z-bounds}
    \frac{2}{K+1}\leq z\leq 1\,.
\end{equation}

In Fig. \ref{fig:criticalPointExamples}, we show some simulation examples for different
$z$ and $S$. One can check that all the values of $z$ satisfy Eqs. \eqref{eq:zCondition} and \eqref{eq:z-bounds}.
We see that the possible rational values of $z$, hence the possible cluster patterns, indeed
depend sensitively on the specific integer $S$.
Note that while in principle, at $K = 4$ there are solutions with any $z$ (satisfying Eqs. \eqref{eq:zCondition}) in the range $0.4 < z < 1.0$, the solutions in Fig.~\ref{fig:criticalPointExamples}, which are sampled with random initial conditions, all have $z \sim 2/3$.  This is likely a combination of entropic factors and a biased sampling of solutions with higher total biomass, similar to that observed in \cite{TaylorMultistability}.
From Appendix \ref{sec:alpha1/2analysis}, the total biomass for a given $z$ for general $S,K$ is
\begin{equation}
\label{eq:alpha1/2biomass}
    -E^*=\sum_i N_i^*=\frac{Sz(2-z)}{2+(K-1)z}\,,
\end{equation}
which reaches maximum near (recall that $z$ has to be rational
at any finite $S$)
\begin{equation}
    z\simeq \frac{2}{\sqrt{K}+1}\,.
\end{equation}
This matches with the simulation results for $K=4$.

Note that the choice of $z$ does not fully fix the cluster pattern up to translations, 
since as seen in Appendix \ref{sec:alpha1/2analysis}
there is still one more free parameter determining the boundary abundances of 
the clusters. On the other hand, the total abundance or the value of the Lyapunov 
function $E$ is fixed by $z$ as in Eq. \eqref{eq:alpha1/2biomass}.
Therefore, the set of local minima of $E$ at 
$\alpha=1/2$ are highly degenerate and characterized by a rational number $z$ 
only. Such structure of commensurate phases \cite{CommensuratePhases} at the ground states is known in
statistical mechanics. An example is the 1D Ising model with long-range 
antiferromagnetic interactions \cite{DevilStaircase}, in which the ground states
are characterized by rational magnetization.
On the other hand, if neglecting the free parameter for boundary abundances, here the total number of stable fixed points
drops significantly and becomes polynomial in $S$ instead. Counting the solutions of
Eq. \eqref{eq:zCondition} and their corresponding translations of patterns, we see that the 
number of stable fixed points grows asymptotically as $O(S^2)$.




The phase transition can also be understood in terms of symmetries of the community.
The interaction network has a translational symmetry $i\rightarrow i+1$ in species space,
while as we have seen so far, the final states of the community away from the critical 
point generically break the translational symmetry completely. 
On the other hand, at the critical point the cluster patterns retain some part of the 
translational symmetry in terms of the uniform patterns in gap sizes and maximum 
abundance across clusters. Such symmetry enhancement is the hallmark of phase transitions.

\subsection{Cascade of phase transitions}
\label{subsec:Cascade}


In the interacting phase at $\alpha<1/2$, the possible values of $N_i^*$ are more dense
as shown in Fig. \ref{fig:chains}(b). Nevertheless, the average sizes of gaps 
follow the same qualitative behavior as in the non-interacting phase as we decrease $\alpha$, 
with phase transitions accumulating at another critical point at $\alpha<1/2$; see
Fig. \ref{fig:phasediagram}(b). Moreover, this type of pattern in average sizes
repeats and leads to $K$ critical points in total, corresponding to the domination 
of community-size chains of clusters with the $K$ possible gap sizes $0\leq d\leq K-1$ respectively. 
The cascade of phase transitions ends at $d=0$, which is the phase with a unique fixed point where all species
coexist with abundance $N_i^*=1/(1+2K\alpha)$ for all $i$.
In general, we start at the critical point $\alpha_c^{d=k}$, 
where all clusters
are separated by $d=k$. As we decrease $\alpha$, chains of clusters with $d=k-1$ start
growing, leading to jumps in the average sizes of gaps and clusters. The growth continues
till $\alpha$ reaches another critical point $\alpha_c^{d=k-1}$. Depending on the 
initial conditions
and/or finite size effects for small $S$, community-size chains with $d=k-1$ may appear at $\alpha$ slightly 
above the critical point, but at the critical point such chains appear
in all cases.



In Appendix \ref{sec:criticalPoints}, we 
study particular types of cluster patterns and derive that the critical
points with gap size $d$ are given by
\begin{equation}
\label{eq:criticalPoints}
    \alpha_c^d=\left(1+\frac{1}{\gamma(d,K)}\csc\frac{\pi}{2(2(K-d)+1)}\right)^{-1}\,,
\end{equation}
where $\gamma(d,K)$ is an $O(1)$ coefficient. In Appendix \ref{subsec:upperCriticalPoints}, we show that
\begin{equation}
    \gamma(d,K)=2\,,\,{\rm for}\,K/2\leq d<K\,.
\end{equation}
For $d<K/2$, the exact value of $\gamma(d,K)$ is unknown, but
we observe in Appendix \ref{subsec:lowerCriticalPoints} that
\begin{equation}
    2\leq\gamma(d,K)\lesssim 3\,.
\end{equation}
The transition to the unique fixed point occurs when the full interaction matrix 
$J_{ij}$ becomes positive definite and the Lyapunov
function becomes globally convex. As shown in Appendix \ref{subsec:coexistence}, the critical $\alpha$ is
\begin{equation}
    \alpha_c^{d=0}\simeq \left(1+\csc\frac{3\pi}{2(2K+1)}\right)^{-1}\,,
\end{equation}
which corresponds to $\gamma(0,K)\simeq 3$ for large $K$.
Note that the critical points
behave asymptotically as $\alpha_c^d\sim 1/(K-d)$ for large $K$ and small $d$. The fixed point is unique when
$\alpha<O(1/K)$ for large $K$, which is precisely when mean-field theory applies. Although we can derive the exact value of $\gamma(d,K)$ when 
$K/2 \leq d < K$, the other cases are harder to analyze generally since the 
clusters have sizes 
exceeding $K+1$ and are no longer fully-connected in the interaction network. Nevertheless,
$\gamma(d,K)$ should interpolate between $2$ and $3$ as shown in Appendix \ref{subsec:lowerCriticalPoints}. We also note that the $\alpha_c$ for the first 
$\lfloor K/2\rfloor$ critical points, corresponding to $1\leq K-d\leq \lfloor K/2\rfloor$,
are independent of $K$. For example, the first critical point is always at
$\alpha_c^{d=K-1}=1/2$, and the second critical point is at $\alpha_c^{d=K-2}=(3-\sqrt 5)/2\simeq 0.381$ for $K\geq 4$.

The properties of these critical points are qualitatively similar to the one
at $\alpha=1/2$. Near each critical point, a dense set of phase transitions lead to
patterns with a fixed gap size.
As in the transition at $\alpha = 1/2$, this appears to be a sharp
transition in which the slope of the average gap size as a function of
$\alpha$ diverges as $\alpha$ approaches the critical point from above.
In addition, the local minima of the Lyapunov function at each critical point are
highly degenerate. In Appendix \ref{subsec:upperCriticalPoints}, we
show that the local minima are also characterized by $\max(N_i^*)$ for
$d\geq  K/2$. Away from the critical points, since all the clusters are now always
interacting with each other when $\alpha<1/2$, we expect that the total number of stable fixed 
points is still polynomial in $S$, similar to the case of
$\alpha=1/2$.
As $\alpha$ decreases, the number of stable
fixed points generally continues to decrease for fixed $S,K$ as interactions become
 weaker; this general trend is confirmed at small $S$ by simply
enumerating all possible solutions.

\section{Discussion}
\label{sec:Discussion}

In this paper, we have studied a minimal model of self-organized clusters
based on Lotka-Volterra competition dynamics in ecological communities. 
These clusters arise as the steady states of the communities regardless of the
initial condition. With
only three parameters, namely the number of species, the number of interacting
neighbors, and the interspecific competition strength, we have found a
large set of cluster patterns and a rich phase diagram. We have classified 
the possible cluster patterns in many of the phases, in terms of cluster sizes
and gap sizes between clusters. We have also investigated various phase 
transitions, including a transition between states with and without clusters,
and a cascade of transitions in which dense sets of phase transitions accumulate
at  critical points. Using tools from statistical mechanics, we
have exactly solved the model with nearest neighbor interactions and studied
its critical behavior. 

Although this model is minimal with only three parameters, it captures the essential
mechanism of cluster formation in ecological communities: the interplay between
effects from neutral and niche theories. This mechanism is also similar to the
competition between short-range and long-range interactions in the context of 
pattern formation \cite{seul1995domain,meinhardt2000pattern}.
Moreover, generalized Lotka-Volterra models serve as effective models for many more
complicated ecological models such as consumer-resource models \cite{MacArthurPacking,chesson1990macarthur}, when there are
ecological steady states. Therefore, we expect that the qualitative features of our
model, such as the qualitative form of the phase diagram, are robust under inclusion
of more complicated model detail. Compared to existing similar models in the niche axis literature
\cite{SchefferCluster, SchefferClusterExplained}, our model also manages to produce stable instead of transient cluster patterns.


Even within our model, there are assumptions for mathematical simplicity that one 
can easily relax without significantly changing the qualitative results.
First, one may consider
a non-periodic, open boundary condition instead of the periodic boundary condition we have imposed. The adjacency matrix has the same form 
as in Eq. \eqref{eq:adjacencyMatrix} but with $d(i,j)=|i-j|$. In other words,
interactions beyond the species boundary ($i<1$ or $i>S$) are cut off instead of
being wrapped to the other boundary. Since all the interactions are local,
we expect that the boundary condition should impact very little on the 
cluster patterns; see \cite{TaylorMultistability} for such an analysis in
the niche axis model. It remains interesting, however, to analyze precisely
the finite size effects due to the boundary condition.

Next, one may replace our banded
adjacency matrix $A_{ij}$ with a smoother, less localized kernel. A common form of smooth kernel is
\begin{equation}
    A_{ij}=
    \exp\left(-\left(\frac{d(i,j)}{K}\right)^\delta\right)\,,
\end{equation}
for $i\neq j$, where $\delta$ is an exponent controlling the extent of 
localization. Note that our model corresponds to the limit 
$\delta\rightarrow\infty$, while the fully-interacting ecosystems without 
clusters correspond
to $\delta\rightarrow 0$. It implies that there are self-organized clusters
only when $\delta$ exceeds some critical value. It is known that for 
$\delta< 2$, there is no pattern formation in the niche axis model 
\cite{NicheAxisTransition}. There is also a phase transition at $\delta=2$
depending on details of the model \cite{NicheAxisTransition,
  NonSmoothKernel, leimar2013limiting}.
We expect that the same is true for self-organized clusters, but an analytical proof remains elusive.

On the other hand, despite the above theoretical robustness of our model, 
we emphasize that our results are still far from being able to fit cluster patterns in realistic
data, hence our idealized model is mainly for conceptual understanding
regarding
the mechanisms 
and phase structures of self-organized clusters.
In more realistic ecosystems, interactions
are heterogeneous across species, and there has been 
some investigation
of using random 
matrices to model these interactions in the large system limit \cite{akjouj2024complex,cui2024houches}.
In the context of 
such
random ecosystems,
there can still be ecological and evolutionary mechanisms
that promote high diversity at the strain level despite limited niche diversity \cite{goyal2022interactions,goyal2025paradox,feng2025theory},
thus leading to some extent of species or strain clustering. Still, these systems are 
disordered and we expect less clustering compared to our model with a regular ecological
network. As a result, we expect that the phase diagram described 
in Sec. \ref{sec:Analysis} is no longer exact in real ecosystems. It remains
interesting to generalize our phase diagram to more random systems, possibly in ways similar to the
statistical mechanics approaches to fully-interacting random ecosystems \cite{BuninGLV,BuninMarginalStability}.



Another key assumption of our model is a one-dimensional niche space for the 
species. The traditional niche axis literature also assumed so (see, however, 
\cite{NicheAxis2D} for a counterexample) since one-dimensional niche spaces are
mathematically more tractable then the higher-dimensional ones, but one-dimensional 
niche spaces are not expected in general ecosystems.
From a statistical mechanics perspective, higher-dimensional niche spaces are also
important since many one-dimensional systems are 
qualitatively distinct from their higher dimensional counterparts due to 
the presence of domain walls. 
It is more subtle, however, to define our model with a discrete niche space in 
higher dimensions, since there are many inequivalent choices for the underlying 
lattice of the niche space, with different point-group symmetries in addition to 
translations. In statistical mechanics, it is well established that the choice of 
lattice connectivity and interactions can qualitatively change the phases, spatial
patterns, and ordering kinetics \cite{moessner2006geometrical,denholm2019topology}, yet there is no reason to expect any 
particular symmetry for the niche space in our context.
Still, we do expect self-organized clusters also in 
higher-dimensional niche spaces based on the intuition of niche and neutral theories
described in Sec. \ref{sec:Introduction}. From numerical simulations, it seems also 
that some features of our phase diagram such as the emergence of long-range 
correlations apply similarly in higher dimensions. The precise mathematical treatment
of these models, as well as possible interplay with physical space, are left for
future work.

\begin{acknowledgments}
  We would like to thank Amer Al-Hiyasat, Guy Bunin,
  Akshit Goyal,
  Pavel Krapivsky, Pankaj Mehta, 
James O'Dwyer, Sidney Redner, and Daniel Swartz for helpful discussions. MK is supported by the NSF through grant DMR-2218849.
The work of WT was supported in part by a grant from the
Schmidt Futures Foundation.
WT would like to thank the Santa Fe Institute, where part of this work was carried out.
This work also made use of resources provided by subMIT
at the MIT physics Department \cite{bendavid2025submitphysicsanalysisfacility}.
\end{acknowledgments}

\appendix

\section{Simulation method}
\label{sec:simulation}

Here we explain the simulation method of our model in Sec. \ref{sec:Model}.
In general, the final state of the community
depends sensitively on the initial condition. To collect examples of fixed points for a given system,
we run simulations of Eq. \eqref{eq:GLV} with random initial abundances
sampled from a uniform distribution between $0$ and $1$, 
until the 
community reaches a steady state. We observe that the simulation may take up to 
$t=10^4\sim10^5$ to reach convergence for the parameter ranges we are interested in; for simplicity we run the simulation until 
$t=10^6$ in all cases. 
To ensure that the community reaches an uninvadable fixed point
in simulations, we add a small positive migration rate $\lambda$ into the 
dynamics, i.e., the right side in Eq. \eqref{eq:GLV}. In other words, the dynamics
becomes
\begin{equation}
    \frac{dN_i}{dt}=N_i\left(1-N_i-\alpha\sum_{j\neq i}A_{ij}N_j\right)+\lambda\,.
\end{equation}
In simulations, we set $\lambda=10^{-10}$ and consider 
a species to be extinct if $N_i^*<10^{-5}$. We set $\lambda=0$ in analytical calculations unless specified.

\section{Motivation for canonical ensemble}
\label{sec:motivation}

In this appendix, we show that the canonical ensemble in Eq. \eqref{eq:canonicalEnsembleProb} can arise in two independent
ways.

\subsection{Demographic noise}

The first way comes from the saddle 
point approximation under the presence of low temperature demographic noise. Although we 
have been studying
deterministic dynamics so far, it is typical in ecology to include demographic noise,
such that the model becomes (assuming It\^o's convention \cite{StochasticMethods})
\begin{equation}
\label{eq:modelWithNoise}
    \frac{dN_i}{dt}=N_i\left(1-N_i-\alpha\sum_{j\neq i}A_{ij}N_j\right)+\sqrt{N_i}\xi_i(t)+\lambda\,,
\end{equation}
Here, the noise $\xi_i(t)$ is a Gaussian process satisfying
\begin{equation}
    \left<\xi_i(t)\right>=0\,,\quad \left<\xi_i(t)\xi_j(t')\right>=T\delta_{ij}\delta(t-t')\,,
\end{equation}
where $T$ is the temperature of the noise; see also \cite{TopologicalGlass} for 
a similar setup for sparse and tree-like interaction networks. 
Nevertheless, if only the demographic noise
is added into the model, every species will become extinct in the 
stationary state. To avoid such global extinction, we must also add a regulator 
such as a constant migration
rate $\lambda>T/2$ into the model. In \cite{BuninMarginalStability,DemographicNoise}, it was shown that the stationary distribution of abundances $N_i$
follow a Boltzmann distribution
\begin{equation}
    p(N_i)\propto e^{-\beta E'}:=e^{-\beta(E-\lambda'\sum_i\log N_i)}\,,
\end{equation}
where $\beta=1/T$ is the inverse temperature of the noise, and 
$\lambda'=2\lambda-T>0$. While the general distribution is hard to analyze due
to continuous and positive degrees of freedom, we can focus on the low 
temperature regime $T\sim\lambda\ll 1$ so that we can apply the saddle point
approximation.
The saddle points satisfy
\begin{subequations}
\begin{align}
    \left.\frac{\partial E'}{\partial N_i}\right|_{N_i=N_i^*}&=-2\left(1-\sum_{ij}J_{ij}N_j^*\right)-\frac{\lambda'}{N_i^*}\\
    &=-2g_i-\frac{\lambda'}{N_i^*}=0\,.
\end{align}
\end{subequations}
The solution can be divided into two cases. The equation can be solved by
$N_i^*\sim O(1)$ satisfying by $g_i=-\lambda'/2N_i^*\simeq 0$, which
corresponds to steady surviving species in the original model. The equation can also
be solved by an $O(1)$ and negative $g_i$ and $N_i^*=-\lambda'/2g_i=O(\lambda')$,
which corresponds to uninvadable extinct species in the original model. 
Therefore, the saddle points correspond to the stable fixed points in the original
model up to $O(\lambda')$ corrections.

Now, the saddle point approximation states that the partition function is
approximated by
\begin{subequations}
\begin{align}
    Z\sim&\sum_\mathrm{saddle}e^{-\beta F^*(N_i^*)}\\
    :=&\sum_\mathrm{saddle}\sqrt{\left(\frac{2\pi}{\beta}\right)^S\left(\det \frac{\partial^2 E'}{\partial N_i\partial N_j}\right)^{-1}}e^{-\beta E'(N_i^*)}\,,
\end{align}
\end{subequations}
where the entropic correction from thermal fluctuations is determined by the 
second derivative
\begin{equation}
    \frac{\partial^2 E'}{\partial N_i\partial N_j}=2J_{ij}+\frac{\lambda'}{N_i^{*2}}\delta_{ij}\,.
\end{equation}
When both $i$ and $j$ are extinct, the above is dominated by the
second term, which is at order $1/\lambda'$. Therefore, the determinant is
dominated by
\begin{equation}
    \det\frac{\partial^2 E'}{\partial N_i\partial N_j}\simeq \left(\det 2J_{ij}^\mathrm{survive}\right)\prod_{\mathrm{extinct}\,i}\frac{4g_i^2}{\lambda'}\,,
\end{equation}
where $J_{ij}^\mathrm{survive}$ is the positive definite submatrix of $J_{ij}$ 
for surviving species. We finally obtain the free energy:
\begin{subequations}
\begin{align}
    F^*\simeq&\,E^*+\Delta F^*\\
    :=&\,E^*-\lambda'\sum_i\log N_i^*-\frac{ST}{2}\log(2\pi T)\\
    &+\frac{T}{2}\log\det 2J_{ij}^\mathrm{survive}-\frac{T}{2}\sum_{\mathrm{extinct}\,i}\log\frac{\lambda'}{4g_i^2}\,.
\end{align}
\end{subequations}

To connect the above approximation to Eq. \eqref{eq:canonicalEnsembleProb}, we have to 
ensure that the fluctuation correction $\Delta F^*$ is much smaller than the free energy
obtained from Eq. \eqref{eq:canonicalEnsembleProb}. First, we notice that although $\Delta F^*$ highly depend on the abundances
in a saddle point, $\Delta F^*$
always scales at most as $O(ST\log T)$, which is a small correction to 
$E^*\sim O(S)$ for $T\ll 1$. Next, in the large $S$ limit, 
the entropic contribution to the free energy from
Eq. \eqref{eq:canonicalEnsembleProb} comes from summing the highly degenerate saddles due to 
the permutations of cluster patterns. The corresponding free energy scales as $O(ST\log S)$, which is also much larger than $\Delta F^*$ when $S\gg 1/T$.
In conclusion, we can approximate the above partition function as
\begin{equation}
    Z\sim \sum_\mathrm{saddle}e^{-\beta E^*}\,,
\end{equation}
i.e., the same as Eq. \eqref{eq:canonicalEnsembleProb} as long as $S\gg 1/T\gg 1$.

\subsection{Random initial conditions}
\label{subsec:randomInitialConditions}

\begin{figure}
    \centering
    \includegraphics[width=1\columnwidth]{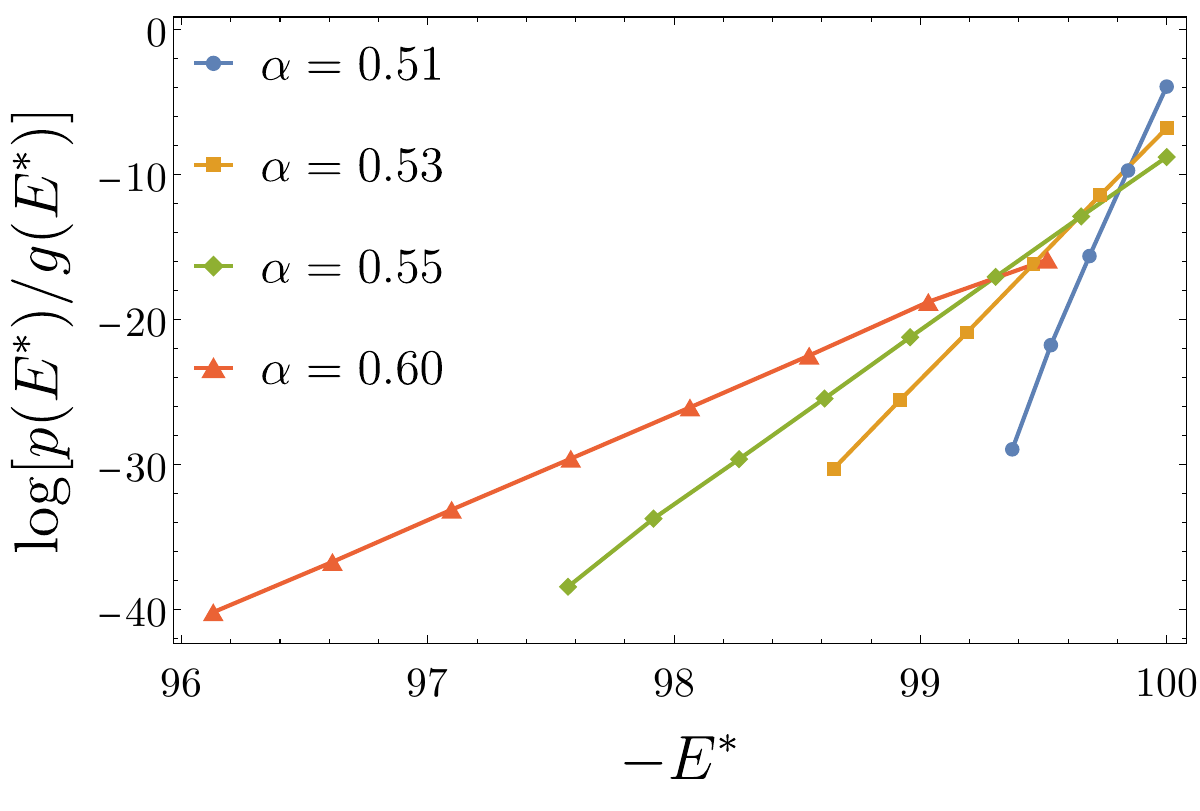}
    \caption{\textbf{Canonical ensemble of steady states from random initial
    conditions.} The probabilities $p(E^*)$ for $S=200,K=1$, and 
    $\alpha=0.51,0.53,0.55,0.60$ are obtained
    from simulations with $10^4$ set of random initial conditions uniformly
    sampled between $0$ and $1$. The degeneracies $g(E^*)$ are from Eq. 
    \eqref{eq:degeneracy}. The linear plots demonstrate the canonical ensemble 
    relation $p(E^*)\propto g(E^*)e^{-\beta E^*}$.}
    \label{fig:randomIC}
\end{figure}

The second independent way to motivate the canonical ensemble assumes deterministic 
dynamics, but considers a  
large ensemble of random initial 
conditions, which evolves to an ensemble of stable fixed points. A good 
example is to sample each initial abundance uniformly between $0$ and $1$,
since the abundances in the stable fixed points are always bounded in this 
range. Using the
locality of the interactions, it was argued in \cite{TaylorMultistability} that such a distribution of
stable fixed points at the final state can be roughly approximated by Eq.
\eqref{eq:canonicalEnsembleProb}.
More precisely, it follows from the Markov chain central limit theorem that the joint distribution of $E^*$ and $\log p$ becomes a bivariate Gaussian distribution
as $S \rightarrow \infty$.
Note that unlike the case with demographic noise, here $\beta$
is
a natural feature of the system instead of being arbitrarily 
chosen.
While $\beta$ depends weakly on the precise choice of the distribution on initial conditions, it is relatively independent of this choice as long as the distribution is invariant under cyclic rotation of species $i \rightarrow i +1$.
To check this approximation for $K=1$, we
fix several values of $\alpha$ and run the simulations with $10^4$ random initial
conditions, in which each initial abundance is sampled uniformly between $0$
and $1$. We count the frequencies of $E^*$ to obtain the probabilities 
$p(E^*)$. Using the relation between $E^*$
and its degeneracy (see also Eq. \eqref{eq:degeneracy}), we can then plot 
Eq. \eqref{eq:canonicalEnsembleProb} as in Fig. \ref{fig:randomIC}. We observe that Eq. 
\eqref{eq:canonicalEnsembleProb} indeed describes the ensemble from random
initial conditions  well.

Note that when connecting Fig. \ref{fig:randomIC} to Eq. 
\eqref{eq:canonicalEnsembleProb}, we have assumed that all
configurations
degenerate 
in $E^*$ have equal probability to one another.
We find in simulations that
this assumption is approximately true for $K=1$. Nonetheless,
for higher $K$ where the set of states is more diverse, the above assumption is no longer true. There would be more
spread/noise in the joint $(- E^*, \log p)$ distribution.

\section{Exact analysis on $K=1$}
\label{sec:moreKequals1}

In this appendix, we exactly solve the model in Sec. \ref{sec:Kequals1} and its statistical mechanics variant with full derivations.
First, we classify the stable and uninvadable cluster patterns using exact
expressions for abundances. Next, we perform the statistical mechanics
analysis using both grand canonical and canonical ensembles, which are equivalent in
the large $S$ limit. While the calculations are less intense in the grand canonical ensemble, it is easier in the
canonical ensemble to extract more sophisticated information such as the full form of two-point correlation function.
Finally, we derive analytical approximations to obtain the correlation length, which is independent of temperature. We discuss its implications in terms of domain walls.

\subsection{Stable and uninvadable configurations}
\label{subsec:uninvadable}

We consider configurations where $K = 1$ and $1/2 <\alpha <1$.
First, stability requires the cluster to have even sizes (if larger than $1$) 
\cite{BuninSparse}. As we will further show in Appendix \ref{subsec:chainProof}, a cluster
with size $n=2l$
is stable if 
\begin{equation}
    \alpha <\frac{1}{2\cos\frac{\pi}{2l+1}}\,.
\end{equation}
Therefore, the stable (but not necessarily uninvadable) cluster sizes are $n=1,2l$ where 
$1\leq l\leq l_\mathrm{max}$ satisfies the above inequality. All clusters are separated by distance $d = 1$. 

We now turn to uninvadability, which requires more information about the abundances.
Let $N^l_1,\cdots,N^l_{2l}$ be the abundances of each 
species in a cluster of size $2l$. The corresponding interaction matrix 
$J_{ij}$ is
\begin{equation}
    J_{ij}=\left(\begin{array}{cccc}
    1 & \alpha\\
    \alpha & 1 & \ddots\\
     & \ddots & \ddots & \alpha\\
     &  & \alpha & 1
    \end{array}\right)\,,
\end{equation}
and the abundances are
\begin{equation}
    N^l_i=\sum_j J_{ij}^{-1}\,.
\end{equation}
By inverting the
tridiagonal matrix, one can show that (Corollary 4.2-4.4 in \cite{TridiagonalInverse})
\begin{subequations}
\begin{align}
    N^l_1=N^l_{2l}&=\frac{1+\alpha\left(J_{11}^{-1}+J_{1(2l)}^{-1}\right)}{1+2\alpha}\\
    &=\frac{1}{1+2\alpha}\left(1+\frac{U_{2l-1}(1/2\alpha)-1}{U_{2l}(1/2\alpha)}\right)\,,\\
    \label{eq:Chebyshev}
    \sum_i N^l_i&=\frac{2l+2\alpha N^l_1}{1+2\alpha}\,,
\end{align}
\end{subequations}
where $U_m(x)$ is the Chebyshev polynomial of the second kind. Note that we can express the abundance sum 
in terms of the boundary abundances $N^l_1=N^l_{2l}$ only. Using the 
definition of Chebyshev polynomials, it is useful to
further rewrite the above as
\begin{subequations}
\begin{align}
    &N_1^l
    =\frac{1}{1+2\alpha}\left(1+\frac{\sin 2l\theta-\sin\theta}{\sin (2l+1)\theta}\right)\\
    =\;&\frac{1}{1+2\alpha}\left(1+\frac{\sin (2l+1)\theta\cos\theta-(1+\cos(2l+1)\theta)\sin\theta}{\sin (2l+1)\theta}\right)\\
    =\;&\frac{1}{1+2\alpha}\left(1+\frac{1}{2\alpha}-\frac{\sin\theta}{\sin(2l+1)\theta/(1+\cos(2l+1)\theta)}\right)\\
    =\;&\frac{1}{2\alpha}-\frac{\sin\theta}{(1+2\alpha)\tan\frac{2l+1}{2}\theta}\,,
\end{align}
\end{subequations}
where $\theta=\cos^{-1}(1/2\alpha)$.
We see that for fixed $\alpha$, $N^l_1$ is 
increasing in $l$ but always less than $1/2\alpha<1$.

Using the above expressions, we can now study the invasion fitness of the 
extinct species between two clusters. First, if both clusters have sizes 
greater than $1$, the invasion fitness is bounded by
\begin{equation}
g_i\geq 1-2\alpha N^{l_\mathrm{max}}_1>1-2\alpha\cdot\frac{1}{2\alpha}=0\,,
\end{equation}
 where $l_\mathrm{max}$ is defined by Eq.~ (\ref{eq:alphaInequality});
hence the gap is invadable and one of the clusters must have size $1$. If both
clusters have size $1$, the invasion fitness is
\begin{equation}
    g_i=1-2\alpha<0\,.
\end{equation}
Alternatively, if the cluster sizes are $1, 2l_\mathrm{max}$, the invasion 
fitness is
\begin{subequations}
\begin{align}
    g_i&=1-\alpha\left(1+N^{l_{\mathrm{max}}}_1\right)\\
    &\leq \frac{1}{2}-\alpha+\frac{\alpha\sin\frac{\pi}{2(l_\mathrm{max}+1)+1}}{(1+2\alpha)\tan\frac{2l_\mathrm{max}+1}{2}\cdot\frac{\pi}{2(l_\mathrm{max}+1)+1}}\\
    &=\frac{1}{2}-\alpha+\frac{\alpha \sin^2\frac{\pi}{2(l_\mathrm{max}+1)+1}}{(1+2\alpha)\cos\frac{\pi}{2(l_\mathrm{max}+1)+1}}\\
    &\leq \frac{1}{2}-\alpha+\frac{\alpha(1-(1/2\alpha)^2)}{(1+2\alpha)/2\alpha}=0\,,
\end{align}
\end{subequations}
hence the gap is uninvadable. Nonetheless, if the second cluster has size $1<n<2l_\mathrm{max}$, a similar calculation yields
\begin{subequations}
\begin{align}
    g_i&\geq 1-\alpha\left(1+N^{l_{\mathrm{max}}-1}_1\right)\\
    &> \frac{1}{2}-\alpha+\frac{\alpha\sin\frac{\pi}{2l_\mathrm{max}+1}}{(1+2\alpha)\tan\frac{2(l_\mathrm{max}-1)+1}{2}\cdot\frac{\pi}{2l_\mathrm{max}+1}}\\
    &=\frac{1}{2}-\alpha+\frac{\alpha \sin^2\frac{\pi}{2l_\mathrm{max}+1}}{(1+2\alpha)\cos\frac{\pi}{2l_\mathrm{max}+1}}\\
    &> \frac{1}{2}-\alpha+\frac{\alpha(1-(1/2\alpha)^2)}{(1+2\alpha)/2\alpha}=0\,,
\end{align}
\end{subequations}
hence the gap is invadable. The above derivation uses the inequality in Eq.
\eqref{eq:alphaInequality} and the facts that $\sin \theta/\tan (2l+1)\theta/2$
is decreasing in $\theta$ for $\theta<2\pi/(2l+1)$, and that $\sin^2\theta/\cos\theta$ is increasing in $\theta$ for $\theta<\pi/2$.

Combining all the results, we conclude that the only stable and uninvadable adjacent 
cluster
patterns are $n=1,1$ and $n=1,2l_\mathrm{max}$. Following Sec.~\ref{sec:Kequals1}, the symbol $l$ now refers to $l_\mathrm{max}$ specifically.
To facilitate the analysis below, we define the
two following cluster patterns according to the rules for the possible stable fixed 
points:
\begin{enumerate}
    \item A surviving species followed by an extinct species, corresponding to $n=1$. 
    This pattern has total abundance $1$ and size $2$.
    \item A group of $2l$ surviving species, followed by an extinct species, a surviving
    species, and an extinct species. This pattern corresponds to $n=2l$, immediately
    followed by $n=1$ since there cannot be two adjacent clusters with $n=2l$. This
    pattern has total abundance $\sum_i N^l_i+1$ and size $2l+3$, where $N^l_1,\cdots,N^l_{2l}$ are the abundances of each species in the large cluster.
\end{enumerate}
We can then obtain the degeneracy $g(E^*)$ for steady states with Lyapunov 
function $E=E^*$. Suppose that the second cluster 
pattern occurs $b_{2l+3}$ times. We see that the number of times the first pattern arises is
\begin{equation}
    b_2=\frac{1}{2}\left(S-(2l+3)b_{2l+3}\right)\,.
\end{equation}
We also have
\begin{align}
  -E^*&=b_2+b_{2l+3}\left(\sum_i N^l_i+1\right)\,,
\end{align}
where $\sum_i N_i^l$ is given in Eq. \eqref{eq:Chebyshev}.
We can solve these two linear equations to get $b_2, b_{2 l +3}$ as functions 
of $E^*$.  The degeneracy for a given $E^*$ can then be computed 
combinatorially:
\begin{equation}
    \label{eq:degeneracy}
    g(E^*)=\frac{S}{b_2 (E^*) + b_{2 l +3}(E^*)}\cdot \left(\begin{matrix}b_2(E^*)+b_{2l+3}(E^*)\\
    b_2(E^*)\end{matrix}\right)\,.
\end{equation}
The binomial coefficient counts the combinations of cluster patterns, and the
prefactor counts the cyclic permutations that lead to distinct abundance 
configurations from a cluster pattern.
Note that this equation holds independent of the symmetry structure of the pattern of clusters.

\subsection{Grand canonical ensemble}

We can now solve for the statistics of the ensemble using the transfer matrix 
method. Nevertheless, it is
mathematically harder to work with the canonical ensemble directly, since the
cluster sizes vary while the community size $S$ is fixed. Instead, we allow
$S$ to fluctuate and work with the grand canonical ensemble by introducing a
chemical potential $\mu$. The two ensembles should agree in the thermodynamic
limit $S\rightarrow\infty$. For comparison, we provide the analysis for the
canonical ensemble in the next section.
Since patterns 1, 2 from the previous subsection can appear in any combinations without any correlation 
between patterns, the transfer matrix is simply
\begin{equation}
    M=\left(\begin{matrix}
        e^{\beta(1+2\mu)} & e^{\beta(\sum_i N^l_i+1+(2l+3)\mu)}\\
        e^{\beta(1+2\mu)} & e^{\beta(\sum_i N^l_i+1+(2l+3)\mu)}
    \end{matrix}\right)\,.
\end{equation}
We then obtain the grand partition function:
\begin{equation}
    \mathcal Z=\left(1-e^{\beta(1+2\mu)}-e^{\beta(\sum_i N^l_i+1+(2l+3)\mu)}\right)^{-1}\,.
\end{equation}
We find the chemical potential in the thermodynamic limit by requiring
\begin{subequations}
\begin{align}
    \left<S\right>&=\frac{1}{\beta}\frac{\partial(\log\mathcal Z)}{\partial\mu}\\
    &=\frac{2e^{\beta(1+2\mu)}+(2l+3)e^{\beta(\sum_i N^l_i+1+(2l+3)\mu)}}{\left(1-e^{\beta(1+2\mu)}-e^{\beta(\sum_i N^l_i+1+(2l+3)\mu)}\right)^2}\rightarrow\infty\,.
\end{align}
\end{subequations}
Therefore, the chemical potential satisfies
\begin{equation}
\label{eq:chemicalPotential}
    1-e^{\beta(1+2\mu)}-e^{\beta(\sum_i N^l_i+1+(2l+3)\mu)}=0\,.
\end{equation}

To proceed, we shall focus on the critical behavior near $\alpha=1/2$. Using 
the result in Eq. \eqref{eq:Chebyshev}, we can approximate
\begin{equation}
    \sum_i N_i^l\rightarrow \frac{2l+1}{2}\,,
\end{equation}
which implies that all the cluster patterns become nearly degenerate as 
$\alpha\rightarrow 1/2$. We then expect that the system is independent of 
temperature; see also Sec. \ref{subsec:domainWall}. Indeed, 
Eq. \eqref{eq:chemicalPotential} now becomes
\begin{equation}
\label{eq:equationForX}
    x^{2l+3}-x^{2l+1}-1=0\,,
\end{equation}
where $x=e^{-\beta(\mu+1/2)}$.
In other words, the system now depends on $x$ instead of $\beta$, and $x$ is determined by $l$ only. To find the
precise value of $x$, we order the 
solutions $x_i$ to Eq. \eqref{eq:equationForX} such that $|x_1|\geq|x_2|\geq\cdots$. It
turns out that $x_1$ is the only real solution, so the chemical potential must be given
by $x=x_1$.

To extract the correlation length in this ensemble, we first need the probabilities
of appearance for each cluster pattern. From the transfer matrix, the longer cluster pattern
appears with probability
\begin{equation}
\label{eq:clusterProbability}
    p_l=\frac{e^{\beta(\sum_i N^l_i+1+(2l+3)\mu)}}{e^{\beta(1+2\mu)}+e^{\beta(\sum_i N^l_i+1+(2l+3)\mu)}}\simeq \frac{1}{x_1^{2l+3}}\,.
\end{equation}
The abundance correlation between different species can then be seen as follows. Suppose
we have $N_i^*=1$ for some $i$, i.e., the beginning of the shorter cluster pattern. We
are interested in the conditional expected value of $N_{i+r}^*$, i.e.,
\begin{equation}
    f(r):=\mathbb E(N_{i+r}^*|N_i^*=1)\,.
\end{equation}
We focus on its asymptotic behavior and take $r\gg l$. Note that starting with other 
abundances for $N_i^*$ does not change the asymptotic behavior, since after that 
we must have $N_j^*=1$ for some $j\leq i+2l+2$.

Using Eq. \eqref{eq:clusterProbability}, we can derive a recursion relation for 
$f(r)$. Notice that there are two possibilities for the abundance $N_{i+2}^*$:
\begin{enumerate}
    \item $N_{i+2}^*=1$ with probability $1-p_l$. If this case is true, the expected
    value then reduces to $f(r-2)$.
    \item $N_{i+2}^*=N^l_1$, i.e., the first species in the large cluster with size
    $2l$, with probability $p_l$. Then we must have $N_{i+2l+3}^*=1$ and the 
    expected value reduces to $f(r-2l-3)$.
\end{enumerate}
Therefore, we have the recursion relation
\begin{equation}
    f(r)=(1-p_l)f(r-2)+p_lf(r-2l-3)\,.
\end{equation}
The recursion relation can be solved by assuming an ansatz $f(r)\propto y^r$ for some $y$.
The possible values of $y$ satisfy
\begin{equation}
    (x_1y)^{2l+3}-(x_1y)^{2l+1}-1=0\,,
\end{equation}
which is the same as Eq. \eqref{eq:equationForX}, hence the solutions are in the form of 
$y_i=x_i/x_1$. The leading solution is $y_1=1$, which does not decay and corresponds to 
the unconditional 
expected value $\left<N_{i+r}^*\right>$. The correlation comes from the first subleading
solution $y_2=x_2/x_1$, and the correlation length is
\begin{equation}
    \xi=\frac{1}{\log |x_1/x_2|}\,.
\end{equation}

\subsection{Canonical ensemble}
\label{sec:canonicalEnsemble}

To analyze the canonical ensemble instead of the grand canonical ensemble, we write down 
the transfer matrix $M$ from $N_i^*$ to $N_{i+1}^*$ with the appropriate Boltzmann 
factors $e^{\beta N_{i+1}^*}$. We have
\begin{equation}
    M=\left(\begin{array}{cccccc}
    0 & 1\\
    e^{\beta} & 0 & e^{\beta N_{1}^{l}}\\
     &  & 0 & \ddots\\
     &  &  & \ddots & e^{\beta N_{2l}^{l}}\\
     &  &  &  & 0 & 1\\
    e^{\beta} &  &  &  &  & 0
    \end{array}\right)\,,
\end{equation}
where the first two rows are states for a cluster with size 1, and the remaining rows
are states for a cluster with size $2l$. Now, the partition function is simply 
$Z=\mathrm{Tr}\,M^S$. To proceed, we note that the eigenvalues of $M$ 
(denoted as $\lambda$) satisfy
\begin{equation}
\label{eq:eigenvalues}
    \lambda^{2l+3}-e^\beta \lambda^{2l+1}-e^{\beta\left(\sum_i N^l_i +1\right)}=0\,.
\end{equation}
Note that this equation is the same as Eq. \eqref{eq:chemicalPotential} with
$\lambda=e^{-\beta\mu}$.
We order the eigenvalues such that $|\lambda_1|\geq|\lambda_2|\geq\cdots$. We observe that
$\lambda_1$ is positive, while $\lambda_2,\lambda_3$ are complex conjugates with 
negative real parts and smaller magnitude than $\lambda_1$.

We can now calculate the following ensemble-averaged quantities.
As a simple example, we have
\begin{equation}
    \left<N_i^*\right>=\frac{1}{S}\frac{\partial(\log Z)}{\partial\beta}\simeq\frac{\lambda_1^{2l+1}+e^{\beta\sum_i N^l_i}(\sum_i N^l_i +1)}{2\lambda_1^{2l+1}+(2l+3)e^{\beta\sum_i N^l_i}}\,.
\end{equation}
Using the relation between $\lambda_1$ and $\mu$, one can check that the above
indeed agrees with the grand
canonical ensemble result in the thermodynamic limit.

On the other hand, it is not obvious how to get the two-point correlation function directly from the
partition function. Instead we calculate
\begin{equation}
    \left<N_i^* N_{i+r}^*\right>=\frac{1}{Z}\mathrm{Tr}\,\left(NM^r NM^{S-r}\right)\,,
\end{equation}
where $N=\mathrm{diag}(1,0,N^l_1,\cdots,N^l_{2l},0)$ is the diagonal matrix of abundances for
each state. The trace can be calculated by diagonalizing $M=UDU^{-1}$, 
which requires the left and right
eigenvectors of $M$. One can check that for eigenvalue $\lambda$, the right eigenvector
is
\begin{align}
    \vec v=&\left(e^{-\beta}\lambda,e^{-\beta}\lambda^2,e^{\beta(N^l_2+\cdots+N^l_{2l})}\lambda^{-2l},\right.\nonumber\\
    &\left.e^{\beta(N^l_3+\cdots+N^l_{2l})}\lambda^{-2l+1},\cdots,\lambda^{-1},1\right)\,,
\end{align}
and the left eigenvector is
\begin{equation}
    \vec w=\frac{\left(\lambda^{2l+2},\lambda^{2l+1},e^{\beta N^l_1}\lambda^{2l},\cdots,e^{\beta\sum_i N^l_i}\lambda,e^{\beta\sum_i N^l_i}\right)}{2\lambda_1^{2l+1}+(2l+3)e^{\beta\sum_i N^l_i}}\,,
\end{equation}
such that $\vec w_i\cdot \vec v_j=\delta_{ij}$. The two-point function can
then be expressed in terms of the eigenvalues $\lambda_i$ and the 
elements of the matrix $\tilde N=U^{-1}NU$:
\begin{align}
    \tilde N_{ij}&=w_i^TNv_j\nonumber\\
    &=\frac{e^{-\beta}\lambda_i^{2l+2}\lambda_j+e^{\beta\sum_k N^l_k}\sum_k N^l_k(\lambda_i/\lambda_j)^{2l+1-k}}{2\lambda_1^{2l+1}+(2l+3)e^{\beta\sum_k N^l_k}}\,.
\end{align}
We now focus on the limit $1\ll r<S/2$ and calculate the two-point function.
The leading contribution of the two-point function is simply $\left<N_i^*\right>^2$.
The connected two-point function is given by the first subleading
contribution, namely
\begin{equation}
    \left<N_i^* N_{i+r}^*\right>-\left<N_i^*\right>^2\simeq 2\mathrm{Re}\,\tilde N_{12}\tilde N_{21}\left(\frac{\lambda_2}{\lambda_1}\right)^r\,.
\end{equation}
Therefore, the connected two-point function alternates between positive and negative
values. Its magnitude oscillates with an exponentially decaying amplitude.

The correlation length $\xi$ can be extracted from the two-point function.
We have
\begin{equation}
    \xi=\frac{1}{\log|\lambda_1/\lambda_2|}= \frac{1}{\log|x_1/x_2|}\,,
\end{equation}
where $x=e^{-\beta/2}\lambda$. Near the critical point $\alpha=1/2$, $x$
indeed satisfies Eq. \eqref{eq:equationForX} as in the grand canonical ensemble.

\subsection{Correlation length}
\label{subsec:corrLength}


We now derive an analytic approximation to $\xi$ to understand its asymptotic
behavior. Recall that we need to solve the equation
\begin{equation}
    x^{2l+3}-x^{2l+1}-1=0\,,
\end{equation}
and find the first two solutions $x_1,x_2$ with the largest magnitudes.
First we rearrange the above and get
\begin{equation}
    x^{2l+1}(x^2-1)=1\,.
\end{equation}
When $l$ is large, we observe that $|x_1|,|x_2|$ are close to $1$ with $O(1/l)$ corrections, so that
$|x^2-1|$ is small but $|x|^{2l+1}$ is large. Therefore, we expand $x_\pm=\pm 1+\delta x_\pm$ with $\delta x_\pm=O(1/l)$ and get
\begin{equation}
    2\delta x_\pm(1\pm\delta x_\pm)^{2l+1}\simeq 2\delta x_\pm e^{\pm(2l+1)\delta x_\pm}=1\,.
\end{equation}
Solving the above equation yields
\begin{subequations}
\begin{align}
\label{eq:Wfunction}
    x_1&\simeq1+\frac{1}{2l+1}W\left(l+\frac{1}{2}\right)\,,\\
    x_2&\simeq-1-\frac{1}{2l+1}W\left(-l-\frac{1}{2}\right)\,,
\end{align}
\end{subequations}
where $W(x)$ is the principal branch of the Lambert $W$ function. We then have
\begin{equation}
    \xi\simeq \frac{2l+1}{W(l+1/2)-\mathrm{Re}\,W(-l-1/2)}\,.
\end{equation}
Finally, using the relation
\begin{equation}
    \mathrm{Re}\,W(-x)\simeq W(x)-\frac{\pi^2}{2W(x)^2}\,,
\end{equation}
which holds for large positive $x$, we arrive at
\begin{equation}
\label{eq:correlationLengthFinal}
    \xi\simeq \frac{4}{\pi^2} W(l)^2l\simeq \frac{4}{\pi^2}l(\log l)^2\,.
\end{equation}



\subsection{Domain walls}
\label{subsec:domainWall}

The temperature independence near the critical point can be understood in 
terms of domain walls between the two cluster patterns with sizes $n=1,2l$. 
Consider $2l$ consecutive species, which can be in either
a single cluster with $n=2l$, or $l$ clusters with $n=1$. 
Recall that near the critical point, we have $\sum N^l_i\rightarrow (2l+1)/2$, 
which is almost the same as the total abundance of $l$ clusters with $n=1$.
Therefore, both configurations for $2l$ species have almost the same energy near
the critical point. For such a 1D statistical mechanical system, domain walls between
the two configurations should proliferate at any finite $\beta$. As a result, the
distribution of cluster patterns
are dominated by domain wall contributions and become independent of $\beta$.

The same distribution of cluster patterns can also be seen in the transfer 
matrix. Using the expression of $x_1$ in Eq. \eqref{eq:Wfunction},
the probability for longer clusters in Eq. \eqref{eq:clusterProbability} becomes
\begin{equation}
    p_l\simeq 1-\frac{1}{l}\,.
\end{equation}
It means that both cluster patterns occupy almost 
the same number
of species in average near the critical point, which is indeed the expected behavior
when domain walls proliferate.

Since there are no two adjacent size-$2l$ clusters, we should associate the 
domain walls to only one of the boundaries (say, the left boundary) of the
longer clusters. With this definition of domain walls, we see that two domain
walls must be separated by distance of at least $2l+3$, but any configurations
of domain walls satisfying this distance constraint is a valid ground state.
Such structure of domain walls is known in statistical mechanics. An example
is the 1D Ising model with competing interactions,
or to be precise, a short-range ferromagnetic interaction and a long-range antiferromagnetic
interaction, at a critical coupling. \cite{RednerCompetingIsing,KardarHelix}.

\section{Stability analysis of simple cluster patterns}
\label{sec:simpleClusters}

In this appendix, we study the two simplest cluster patterns: a single cluster and a 
pair of interacting clusters, and classify their possibilities using analytical arguments.
We also derive a bound on the critical $\alpha$ for chains of higher lengths.
We focus on the non-interacting regime $1/2<\alpha<1$, where the
species within a chain of clusters 
do not interact with any surviving species outside the chain. 
We will prove that a single cluster
must have size $1\leq n\leq K+1$. We will also prove that the two interacting clusters 
in a pair must have the same size $2\leq n\leq K$ and are separated by $d=K-1$.

The stability of a set of surviving species is determined by their interaction network.
To facilitate the proof, we will need the following two lemmas about the interaction
networks of surviving species which were proven in \cite{BuninSparse}:
\begin{enumerate}
    \item An interaction network is unstable if there is an unstable subnetwork;
    \item An interaction network is feasible and stable if and only if there is no stable
    fixed point where only a proper subset of species in the network survive, and all the
    extinct species are uninvadable.
\end{enumerate}
Note that here we only focus on networks of surviving species, but not the 
uninvadability for extinct species outside the networks.

\subsection{Single cluster}
\label{subsec:single}

Here we show that for $1/2<\alpha<1$, a single cluster not interacting with other clusters
must have size
$1\leq n\leq K+1$, i.e., the cluster must have a fully-connected interaction network. In Sec. \ref{subsec:Onset}, we have already shown that the cluster 
can have size $1\leq n\leq K+1$. It remains to show that these are the only 
possibilities.

Roughly speaking, if $n>K+1$, not every species in the 
cluster competes in the same way, hence some species may become extinct under strong competition.
In other words, there are alternative stable and uninvadable states, which rule out the
cluster state by the second lemma. Nevertheless, writing down these alternative
states for all $n$ and $\alpha$ is a difficult task. Instead, we make use of the fact that
such an interaction network is unstable in most but not all cases when $n>K+1$. We divide into several cases:
\begin{itemize}
    \item $n=K+2$: Consider a stable fixed point where only the two species at the boundary
    of the network survive, and there are $K$ extinct species in between. We see that
    the two surviving species have $N_i^*=1$ and all the extinct species are uninvadable when $\alpha>1/2$.
    Hence $n=K+2$ is ruled out by the second lemma.
    \item $n\geq K+3,K\geq 6$: We first show that for $K\geq 6$, the network $n=K+3$ is 
    unstable, then the cases $n>K+3$ are implied using the first lemma. Consider the 
    most extreme case $\alpha=1/2$. For $n=K+3$, we have
    \begin{equation}
    2J_{ij}=\left(\begin{array}{cccc|cc}
    2 & 1 & \cdots & 1 & 0 & 0\\
    1 & 2 & \ddots & \vdots & 1 & 0\\
    \vdots & \ddots & \ddots & 1 & \vdots & 1\\
    1 & \cdots & 1 & 2 & 1 & \vdots\\
    \hline
    0 & 1 & \cdots & 1 & 2 & 1\\
    0 & 0 & 1 & \cdots & 1 & 2
    \end{array}\right)\,.
    \end{equation}
    By splitting the matrix into block matrices as indicated above and using the formula
    \begin{equation}
        \det\left(\begin{matrix}A&B\\C&D\end{matrix}\right)=\det A\det\left(D-CA^{-1}B\right)\,,
    \end{equation}
    when the submatrix $A$ is invertible, we get
    \begin{equation}
        \det\left(2J_{ij}\right)=6-K\leq 0\,.
    \end{equation}
    Therefore, the interaction matrix must have nonpositive eigenvalues, which become
    even lower for $\alpha>1/2$. We then see that the network $n\geq K+3$ is unstable for
    $K\geq 6$.
    \item $K<6$: One can verify that the networks with sizes $K+2\leq n\leq 8$ are all 
    infeasible, and the network $n=9$ (hence $n>9$ by the first lemma) is unstable.
\end{itemize}
Combining the three cases, we have shown for any $K$ that a single cluster 
cannot have size $n\geq K+2$.

\subsection{Pair of clusters}
\label{subsec:pair}

Here we show that for $1/2<\alpha<1$, a pair of interacting clusters must have the same size $2\leq n\leq K$.
and separated by $d=K-1$. By the proof for single clusters in the previous section, it
suffices to consider clusters with size $1\leq n\leq K+1$.
To facilitate the proof, we first prove two results for the case of two 
non-interacting clusters $C_1,C_2$ with sizes $n_1,n_2$
and separated by $d=K$. When $n_1,n_2\leq K$ (instead of $K+1$), we have the
following results regarding uninvadability:
\begin{itemize}
    \item If $n_1=n_2=n$, we have $N_i^*=1/(1+(n-1)\alpha)$ for all surviving 
    species $i$. It can be checked that each extinct species interacts with
    at least $n+1$ surviving species. The invasion fitness for all extinct species satisfies
    \begin{equation}
        g_i=1-\sum_{j}J_{ij}N_j^*\leq 1-\frac{(n+1)\alpha}{1+(n-1)\alpha}< 0\,,
    \end{equation}
    for $\alpha> 1/2$. Therefore, none of the extinct species can invade.
    \item If $n_1<n_2$, we consider the first $n_1$ extinct species closest to $C_2$. Let the minimum distance from an extinct species to the boundary 
    of $C_2$ be $d'$. For $1\leq d'\leq n_1$, the invasion fitness satisfies
    \begin{subequations}
    \begin{align}
        g_i&=1-\sum_{j}J_{ij}N_j^*\\&\leq 1-\frac{(n_2+1-d')\alpha}{1+(n_2-1)\alpha}-\frac{d'\alpha}{1+(n_1-1)\alpha}\nonumber\\
        &\leq 1-\frac{(n_2+1)\alpha}{1+(n_2-1)\alpha}< 0\,,
    \end{align}
    \end{subequations}
    for $\alpha> 1/2$. Therefore, these $n_1$ species cannot invade.
\end{itemize}

Now we turn to a pair of interacting clusters $C_1,C_2$. We assume that 
$C_1,C_2$ do not interact with any other clusters.
First, we notice that for a gap size $d$ between $C_1,C_2$, both clusters
must have size $n>K-d$. Suppose $C_1$ has size $n\leq K-d$.
Consider the surviving species $s$ in $C_2$ that is the closest to $C_1$, and the 
extinct species $e$ next to $s$. Both species interact with all the species in $C_1$. Therefore, the invasion fitness of $e$ satisfies
\begin{subequations}
\begin{align}
    g_e&=1-\alpha\sum_{i\in C_1}N_i-\alpha\sum_{\substack{i\in C_2\\d(i,e)\leq K}}N_i\\
    &>1-\alpha\sum_{i\in C_1}N_i-N_s-\alpha\sum_{\substack{i\in C_2\\d(i,e)\leq K+1}}N_i=g_s=0\,.
\end{align}
\end{subequations}
We see that the extinct species can invade the community, which is a 
contradiction.

We can now rule out various interacting pairs by removing species that are the closest to 
the gap, and applying the second lemma.
Suppose $C_1,C_2$ have sizes $K-d<n_1,n_2\leq K+1$, and are separated by a general gap size $d<K$.
We then have the 
following results:
\begin{itemize}
    \item If $n_1=n_2=n$, we can remove species symmetrically until the gap size becomes $K-1$
    or $K$. The first case is our desired cluster pattern, and the second case is 
    uninvadable. Note that the clusters still have nonzero sizes after the
    removal since $n\geq K-d$.
    \item If $n_1<n_2$, we can remove species in $C_2$ until the two cluster
    sizes become equal, or the gap size becomes $K$. The first case reduces back to the 
    case of $n_1=n_2$. The second case is uninvadable since we have removed $K-d<n_1$ 
    species from $C_2$.
\end{itemize}
Therefore, a pair of interacting clusters must have the same size and be separated by $d=K-1$.


Finally, we determine the possible cluster size $n$, which depends on $\alpha$. First, by
symmetry the abundances must have the pattern
\begin{equation}
    N_i^*=x,\cdots,x,y,0,\cdots,0,y,x,\cdots x\,.
\end{equation}
All the surviving species separated from the gap have the same abundance $x$ since they 
only interact with their whole cluster.
By solving the conditions for fixed points, we get
\begin{equation}
    x=\frac{1}{1+(n-1)\alpha-\alpha^2}\,,\quad y=\frac{1-\alpha}{1+(n-1)\alpha-\alpha^2}\,,
\end{equation}
which are both positive. For $2\leq n\leq K$, the invasion fitness of the extinct species 
satisfy
\begin{align}
    g_i=1-\sum_{j}J_{ij}N_j^*&\leq 1-\alpha(nx+2y)\nonumber\\
    &=\frac{1-3\alpha+\alpha^2}{1+(n-1)\alpha-\alpha^2}<0\,,
\end{align}
for $\alpha>1/2$. Therefore, the cases $2\leq n\leq K$ are feasible and uninvadable. 
On the other hand, one can similarly check that the case $n=K+1$ is invadable. At last,
we derive the conditions for stability. Instead of directly computing the eigenvalues
of $J_{ij}$, we can apply the second lemma by removing one species and checking 
invadability. Consider two non-interacting clusters with sizes $n-1$ and $n$. The
invasion fitness of the extinct species next to the smaller cluster is
\begin{equation}
    g_i=1-\frac{(n-1)\alpha}{1+(n-2)\alpha}-\frac{\alpha}{1+(n-1)\alpha}>0\,,
\end{equation}
Hence, the pair of interacting clusters is stable, when
\begin{equation}
    \alpha<\alpha_{n,l=2}=\frac{\sqrt{n+3}+\sqrt{n-1}}{\sqrt{n+3}+3\sqrt{n-1}}\,.
\end{equation}
Note that $1>\alpha_{n,l=2}>1/2$, so the cases $2\leq n\leq K$ are all stable for some range
of $\alpha>1/2$. This concludes our proof.

\subsection{Chain of clusters}
\label{subsec:chainProof}

In principle, we can also classify patterns for chains of interacting clusters with length
$l\geq 3$ using the above techniques, but the process quickly becomes too tedious. Instead,
here we prove an upper bound for the critical $\alpha$ for a chain to become 
stable. Using similar arguments as in the previous section, one can show that 
within an interacting chain with $d=K-1$, the clusters must
have sizes $2\leq n\leq K+1$. In particular, by taking the boundary species of each 
cluster, we see that there is an interaction subnetwork where all clusters have size $2$
and are still separated by $d=K-1$. (When $K=1$, this pattern corresponds to a large cluster with size $n=2l$.) The interaction matrix for this subnetwork has size
$2l$ and is given by
\begin{equation}
\label{eq:bandInteraction}
    J_{ij}=\left(\begin{array}{cccc}
    1 & \alpha\\
    \alpha & 1 & \ddots\\
     & \ddots & \ddots & \alpha\\
     &  & \alpha & 1
    \end{array}\right)\,,
\end{equation}
which has eigenvalues
\begin{equation}
    \lambda_k=1-2\alpha\cos\frac{k\pi}{2l+1}\,,\quad k=1,2,\cdots,2l\,.
\end{equation}
Now by the first lemma, a chain of length $l$ is stable only if the above interaction 
network is stable, which happens when the minimum eigenvalue $\lambda_1$ is positive. 
Therefore, the critical $\alpha$ for a chain of length $l$ to be stable, denoted as 
$\alpha_l$, must satisfy
\begin{equation}
    \alpha_l\leq \frac{1}{2\cos\frac{\pi}{2l+1}}\,.
\end{equation}

\section{Cluster patterns at $\alpha=1/2$}
\label{sec:alpha1/2analysis}
In this appendix, we give full derivations to classify the possible cluster
patterns at the critical point $\alpha=1/2$.

\subsection{Cluster sizes and abundances}

We first find constraints on the cluster sizes and abundances.
Consider a cluster interacting with its two neighboring clusters with $d=K-1$. 
The abundances then have the following pattern:
\begin{equation}
    N_i^* = \cdots,y,0,\cdots,0,x,z,\cdots,z,y',0,\cdots,0,x',\cdots\,,
\end{equation}
where $x,y'$ are the boundary of the cluster, and $y,x'$ belong to the neighboring 
clusters. All the species in the interior of the cluster should have the same abundance
$z$ since they only interact with their whole cluster. Note that when the cluster 
size is $n=2$, the parameter $z$ becomes only a redundant but useful parameter 
instead of real species abundances.
In principle, $x$ or $y$ may vanish and change the gap size; as we will 
see, such solutions have zero measure in the set of all possible solutions.

We let $m=n-2$ be the number of species
in the interior of the cluster. The conditions for fixed points
are
\begin{align}
    x+\frac{1}{2}(y+y'+mz)&=1\,,\nonumber\\
    z+\frac{1}{2}(x+y'+(m-1)z)&=1\,,\\
    y'+\frac{1}{2}(x+x'+mz)&=1\,.\nonumber
\end{align}
We then get the following relations:
\begin{align}
\label{eq:transfer}
    z&=x+y=x'+y'\,,\nonumber\\
    x'&=x-2+(m+2)z\,,\\
    y'&=y+2-(m+2)z\,.\nonumber
\end{align}
Importantly, the first relation implies that the species in the interior of neighboring 
clusters also have abundances $z$. By repeatedly applying the same equation, we see that
all clusters in the community with size $n\geq 3$ have the same interior and maximum abundance 
$z=\max(N_i^*)$.

Using the relations for $x',y'$ in Eq. \eqref{eq:transfer}, we see that generically there is a unique choice of $m$ (for each cluster)
such that both $x',y'$ are positive. We can write alternatively
\begin{equation}
\label{eq:torusMap}
    x'=(x-2)\mod z\,,\quad y'=(y+2)\mod z\,.
\end{equation}
As a result, the abundances at the boundaries vary linearly and wrap around between $0$ and $z$. In other
words, the boundary abundances follow a linear flow on a torus, with size $z\times z$
on the $xy$-plane. The size of each cluster
fluctuates and depends on the number of times the boundary abundances wrap around at each
step of the flow. Note that to satisfy the periodic boundary condition in the interaction network of finite size $S$, the flow must return to its initial value after some number of steps $p$. We 
then have
\begin{equation}
    2p=qz\Rightarrow z=\frac{2p}{q}\,,
\end{equation}
for positive integers $p,q$. Therefore, $z$ must be a rational number.
In addition, the parameter $z$ must satisfy the following necessary condition.
From the above analysis, we note that $p$ is the number of gaps with fixed size 
$K-1$, while $q$ is the total number of times the boundary abundances wrap around 
the interval of length $z$, which is also the total number of surviving species by
Eq. \eqref{eq:transfer}. Therefore, we must have
\begin{equation}
    (K-1)p+q=S\,,
\end{equation}
and the period of the pattern is $S/\mathrm{gcd}(p,q)$. We also see that the total abundance is $\sum_i N_i^*=(q-p)z$.
Note that, however, these
conditions are no longer necessary in the case of strictly infinite $S$, where there are solutions for real $z \leq 1$.

\subsection{Stability and uninvadability}
\label{subsec:alpha1/2proof}

Here we study the stability and uninvadability of the patterns in 
the previous section to find out the constraints on the parameters $z,x,y$.
To begin, we note from Eq. \eqref{eq:transfer} that all clusters have equal size $n$
when $z=2/n$, while the sizes flucutuate between $n$ and $n+1$ when $2/(n+1)<z<2/n$.
In particular, if we require the cluster sizes to satisfy $2 \leq n \leq K +1$ as in the previous subsections, we immediately get the bound
\begin{equation}
 \frac{2}{K +1} \leq z \leq 1 \,,
\end{equation}
which is the same as Eq. \eqref{eq:z-bounds}.

Now we shall show that uninvadability holds for any nonzero $x,y$ when Eq. \eqref{eq:z-bounds} is true.
We consider two adjacent clusters with abundances
\begin{equation}
    N_i^*=x,z,\cdots,z,y',0,\cdots,0,x',z\cdots,z,y''\,,
\end{equation}
where the first cluster has size $n$ and the second one has size $n'\geq n$. 
First consider $n'=2$ where all the $z$'s are absent. Using Eq. \eqref{eq:transfer}, 
the invasion fitness of any extinct species in the gap is
\begin{equation}
    g_i=1-\frac{1}{2}(x+y'+x'+y'')=z-1\,,
\end{equation}
which is nonpositive when $z\leq1$. Now consider $n'>2$ and $n\leq K$. If $n<n'$, the 
extinct species next to the left cluster has the highest invasion fitness. If 
$n=n'$, the extinct species next to the right cluster may also have the highest 
fitness, depending on the boundary abundances. In any case, however, the invasion 
fitness of any extinct species in the gap satisfies
\begin{subequations}
\begin{align}
    g_i&\leq 1-\frac{1}{2}(\min(x+(n-2)z,y''+(n'-2)z)+y'+x'+z)\\
    &=1-\frac{1}{2}(\min(x',y')+2-z+y'+x')<0\,.
\end{align}
\end{subequations}
In contrast, if $n>K$,
the invasion fitness of any extinct species in the gap is
\begin{equation}
    g_i=1-\frac{1}{2}((K-1)z+x'+y'+z)=1-\frac{K+1}{2}z\,,
\end{equation}
which is nonpositive only when $z\geq 2/(K+1)$. Note that $g_i$ vanishes when 
$z=2/(K+1)$ or $z=1$; in these cases additional free parameters for abundances
arise and there is an additional stable fixed point with no extinct
species.
Combining all three cases, we see that 
all the extinct species are indeed uninvadable when Eq. \eqref{eq:z-bounds} holds.

Next, we show that all the cluster patterns derived from Eq. \eqref{eq:transfer}
satisfying $2\leq n\leq K+1$ are stable.
Since all the clusters have fully-connected interaction networks, by the first lemma 
in Appendix \ref{sec:simpleClusters}, it suffices to show the 
stability of a closed chain of $l$ clusters with equal size $n=K+1$. 
Since the interaction matrix has translational symmetry of period $K+1$, we can write
its eigenvectors as
\begin{equation}
    \left(\begin{matrix}\vec v & e^{ik}\vec v & \cdots & e^{(l-1)ik}\vec v\end{matrix}\right)^T\,,
\end{equation}
where $\vec v$ is a $(K+1)$-dimensional vector and $k$ is multiple of $2\pi/l$. 
By permutation symmetry, it suffices to focus on the average of $v_2,\cdots,v_K$, denoted by $\bar v$,
instead of individual components. The eigenvalue equation then 
reduces to
\begin{equation}
    \left(\begin{matrix}
    2 & K-1 & 1+e^{-ik}\\
    1 & K & 1\\
    1+e^{ik} & K-1 & 2
    \end{matrix}\right)\left(\begin{matrix}v_1\\\bar v\\v_{K+1}\end{matrix}\right)=\lambda(k) \left(\begin{matrix}v_1\\\bar v\\v_{K+1}\end{matrix}\right)\,.
\end{equation}
Hence,
\begin{equation}
    \lambda^3-(K+4)\lambda^2+2(K+2-\cos k)\lambda-2(1-\cos k)=0\,.
\end{equation}
By the Routh–Hurwitz criterion, we then see that the eigenvalues are all positive
for all $k$ except $k=0$. For $k=0$, there is a zero eigenvalue and the other two
eigenvalues are positive. Since we already know that the zero eigenvalue corresponds
to the degree of freedom in $x,y$ but not $z$ or the cluster pattern itself, we 
conclude that all the cluster patterns obtained from Eqs. \eqref{eq:transfer} and \eqref{eq:z-bounds} are stable.

To summarize, the stable fixed points at $\alpha=1/2$ are fully characterized by
Eqs. \eqref{eq:zCondition} and \eqref{eq:z-bounds}. For any given $K, S$, we have
solutions for all integer values of $p$ in the range $S/2 K \leq p \leq S/(K
+1)$, each with a multiplicity of $S/\mathrm{gcd}(p,q)$, except when equality holds and additional stable fixed points with no extinction arise.


\section{Positions of critical points}
\label{sec:criticalPoints}

In this appendix, we obtain the critical values of $\alpha$, denoted as 
$\alpha_c$, for accumulation of phase transitions. We derive analytical 
expressions for $\alpha_c$ in some cases, but rely on
numerical simulations otherwise.

\subsection{The critical points when $d\geq K/2$}
\label{subsec:upperCriticalPoints}

We can derive the exact value of 
$\alpha_c^d$ when $d\geq K/2$, by considering an ansatz of cluster patterns 
similar to those in Sec. \ref{subsec:alpha1/2}. Consider two clusters separated by 
gap size $d$ within a community-size chain. As an ansatz of patterns with large clusters,
let the abundances have the following pattern:
\begin{equation}
    N_i^*=\cdots,z,y_1,\cdots,y_{K-d},0,\cdots,0,x_1,\cdots,x_{K-d},z',\cdots\,.
\end{equation}
The first $K-d$ species at the boundary of a cluster have different abundances from the
interior since these species interact with the other cluster. Note that this ansatz 
requires each cluster to have size $n\geq 2(K-d)+1$. On the other hand, we shall focus
on the simple case where each cluster is fully-connected in the interaction network, which
is true only when $d\geq K/2$. Now (the homogeneous part of) the conditions 
for fixed points imply that
\begin{equation}
\label{eq:consistency}
    \left(\begin{array}{ccc|ccc}
    1 &  &  & \beta & \cdots & \beta\\
     & \ddots &  &  & \ddots & \vdots\\
     &  & 1 &  &  & \beta\\
     \hline
    \beta &  &  & 1\\
    \vdots & \ddots &  &  & \ddots\\
    \beta & \cdots & \beta &  &  & 1
    \end{array}\right)\left(\begin{array}{c}
    x_{1}\\
    \vdots\\
    x_{K-d}\\
    \hline
    y_{1}\\
    \vdots\\
    y_{K-d}
    \end{array}\right)=\left(\begin{array}{c}
    z'\\
    \vdots\\
    z'\\
    \hline
    z\\
    \vdots\\
    z
    \end{array}\right)\,,
\end{equation}
where $\beta=\alpha/(1-\alpha)$. Let the above matrix be $M$. If $M$ is not singular, 
there is a unique solution
for all $z,z'$. On the other hand,
if $M$ is singular, the above equations have solutions only when $z=z'$, hence 
all the clusters again have the same maximum abundance. To also ensure stability, 
the critical point should happen at the lowest $\alpha$ such that $M$ is 
singular. By splitting $M$ into block matrices as indicated above, we get
\begin{equation}
    \det(M)=\det\left(I-\beta^2 L\right)\,,
\end{equation}
where $L$ is a lower-right triangular matrix of size $K-d$:
\begin{equation}
    L=\left(\begin{array}{ccc}
     &  & 1\\
     & \ddots & \vdots\\
    1 & \cdots & 1
    \end{array}\right)\,.
\end{equation}
Therefore if $M$ is singular, $\beta^{-1}=(1-\alpha)/\alpha$ should be the absolute values of the 
eigenvalues of $L$, which are \cite{TriangleEigenvalues}
\begin{equation}
    |\lambda_k|=\frac{1}{2}\csc\frac{(2k-1)\pi}{2\left(2(K-d)+1\right)}\,,\quad k=1,\cdots,K-d\,.
\end{equation}
By choosing the maximum eigenvalue at $k=1$, we arrive at
\begin{equation}
    \alpha_c^{d\geq K/2}=\left(1+\frac{1}{2}\csc\frac{\pi}{2(2(K-d)+1)}\right)^{-1}\,.
\end{equation}
Note that we indeed have $\alpha_c^{d=K-1}=1/2$.

Although the flow of the boundary abundances is more complicated, we expect that $z$
should still satisfy some constraints in terms of rational numbers. 
Using the same set of equations in Eq. \eqref{eq:consistency}, one can also show that the
sum $x_1+y_1+\cdots +x_{K-d}+y_{K-d}$ is always a fixed multiple of $z$ regardless of the
remaining free parameters. Therefore, $z=\max(N_i^*)$ uniquely determines the value of $E$ at each of
these critical points.

\subsection{The critical points when $1\leq d<  K/2$}
\label{subsec:lowerCriticalPoints}

\begin{figure}
    \centering
    \includegraphics[width=1\columnwidth]{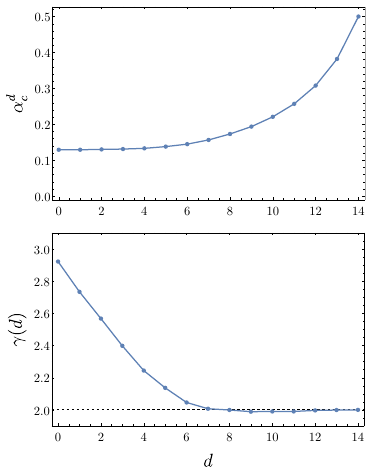}
    \caption{The critical points $\alpha_c^d$ and the corresponding
    parameters $\gamma(d)$ obtained from simulations when $K=15$.}
    \label{fig:critPts}
\end{figure}

In this range of $d$, neither the clusters in simulation results, nor the clusters in the
ansatz in the previous section, are fully-connected in the interaction network. It 
becomes much more difficult to analytically derive $\alpha_c$. Instead, we approximate
the numerical values of $\alpha_c$ using simulations combined with a binary search
algorithm. Then we invert Eq. \eqref{eq:criticalPoints} to find $\gamma(d,K)$. 
Fig. \ref{fig:critPts} shows the resulting $\alpha_c$ for $K=15$. We see that $\gamma(d,K)$ is indeed very close to $2$ when 
$d\geq 8$, and interpolating between $2$ and $3$ otherwise.

\subsection{The critical point at $d=0$}
\label{subsec:coexistence}

Finally, we derive $\alpha_c^{d=0}$, that is the critical point when the community starts to
have a unique coexisting fixed point. The transition happens when the whole interaction
network in Eq. \eqref{eq:adjacencyMatrix} becomes stable. To find the eigenvalues of
$J_{ij}$, we make use of its translation symmetry $i\rightarrow i+1$ to write the 
eigenvectors as
\begin{equation}
    \left(1\quad e^{ik}\quad\cdots\quad e^{(S-1)ik}\right)^T\,,
\end{equation}
where $k$ is multiple of $2\pi/S$. Then, the eigenvalues are
\begin{equation}
    \lambda(k) = 1+2\alpha\sum_{m=1}^K\cos mk=1+\alpha\left(\frac{\sin\left((2K+1)k/2\right)}{\sin\left(k/2\right)}-1\right)\,.
\end{equation}
In the limit of large $S$, we can approximate $k$ as a continuous parameter. The 
minimum eigenvalue is at the minimum positive $k$ such that
\begin{equation}
    \frac{d\lambda}{dk}=0\Rightarrow (2K+1)\tan \frac{k}{2}=\tan \frac{(2K+1)k}{2}\,.
\end{equation}
Although there is no closed-form solution to the above equation in general, we expect
that $(2K+1)k/2$ is close to an odd multiple of $\pi/2$ if $K$ is large. Indeed,
the minimum positive solution can be approximated by $k\simeq 3\pi/(2K+1)$. Now, the coexisting fixed point becomes stable when $\lambda_\mathrm{min}>0$, that is
\begin{equation}
    \alpha_c^{d=0}\simeq \left(1+\csc\frac{3\pi}{2(2K+1)}\right)^{-1}\,.
\end{equation}

\section{Scaling and enumeration of the number of stable solutions}
\label{sec:scaling}

In this appendix, we collect some results on exact enumeration and
scaling properties of the set of stable and uninvadable solutions at
various values of $\alpha$.  As stated in the main text, for any $K$
when $ \alpha >1/2$, the correlation length is finite and the number
of solutions grows exponentially in $S$, going as $e^{\lambda (K, \alpha) S}$ 
at large $S$, while at $\alpha
= 1/2$, there is a phase transition to long-range order and the number
of solutions goes as $S^2$.  For any fixed $K, S$, the number of
solutions in general decreases as $\alpha$ decreases, and the
polynomial scaling at large $S$ continues to decrease for fixed $K$ as
$\alpha$ decreases below 1/2.  Here we give some more quantitative descriptions
of the number of states, focusing in particular on the case $K =
2$.

When $\alpha >1$, a rough estimate of the scaling can be computed by
noting that each isolated species is separated by a gap of size $d$
with $K \leq d
\leq 2 K$ from the next species, with no correlations between adjacent
gap sizes.  Breaking the system into blocks, where each block contains a surviving species and
the subsequent gap, we then have $K +1$ types of blocks. When
averaged over blocks, the mean block size  is $(3 K +2)/2$.  We can then
approximate the number of solutions in a system of size $S$ as
\begin{subequations}
\begin{align}
  {\cal N} (K, S, \alpha > 1)& \sim (K +1)^{2 S/(3 K +2)}\\
  & \simeq e^{\lambda (K, \alpha) S}\,,\\
  \lambda (K, \alpha > 1)& \simeq \frac{2}{3 K +2} \log (K +1)\,.
\label{eq:scaling-estimate}
\end{align}
\end{subequations}
Note, however, that this estimate is slightly off since if we fix the
system size $S$, there are disproportionate numbers of smaller blocks
(for example, if $S = 15$ and we can only have blocks of size 3 or 5,
then we could have 3 blocks of size 5 but 5 blocks of size 3.)

A more precise estimate can be achieved by looking at a
Markov-chain-style
transfer
matrix based on individual species, considering all possible species
positions in each type of block as different states.  For example, for
$K = 2$, we can have blocks of size 3, 4, 5, so we have 12 states
which could be labeled as $3_1, 3_2, 3_3, 4_1, \cdots, 5_5$, where for example $3_1$ refers to the state at a given species that is the first (left-most) position in a block of size 3. We can then define
a transfer matrix $T_{s s'}$ on the set of such states, where $T_{ss'}=1$ only when a
transition from a state $s$ to a subsequent state $s'$ at the next
species to the right is possible, otherwise $T_{ss'}=0$.  For example, $T_{3_1
  3_2}=1$ would be the only nonzero entry for $s = 3_1$, but we would have
$T_{ 3_3 s'}=1$ for $s' = 3_1, 4_1, 5_1$, since the left-most state in any block is possible
following the last state in the previous block.  The total number of states in a
periodic system of size $S$ is then ${\rm Tr} \, T^S$, which is dominated at
large $S$ by
the largest eigenvalue of $T$. 
This is similar to the canonical ensemble method 
used in the $K = 1$ analysis in Appendix \ref{sec:canonicalEnsemble}, and gives a well-defined and systematic
way of computing numbers of states and partition functions for such
systems.

Using this approach, it is
straightforward to compute that, for example, for $K = 2$, the number
of possible states at size $S$ goes as
\begin{equation}
 {\cal N} (2, S, \alpha >1) \sim 1.32472^S \simeq e^{0.281 S}\,,
\label{eq:}
\end{equation}
while the rough estimate in Eq. \eqref{eq:scaling-estimate} gives
$1.31607^S$, i.e., $\lambda \simeq 0.275$.

\begin{figure}
    \centering
    \includegraphics[width=1\columnwidth]{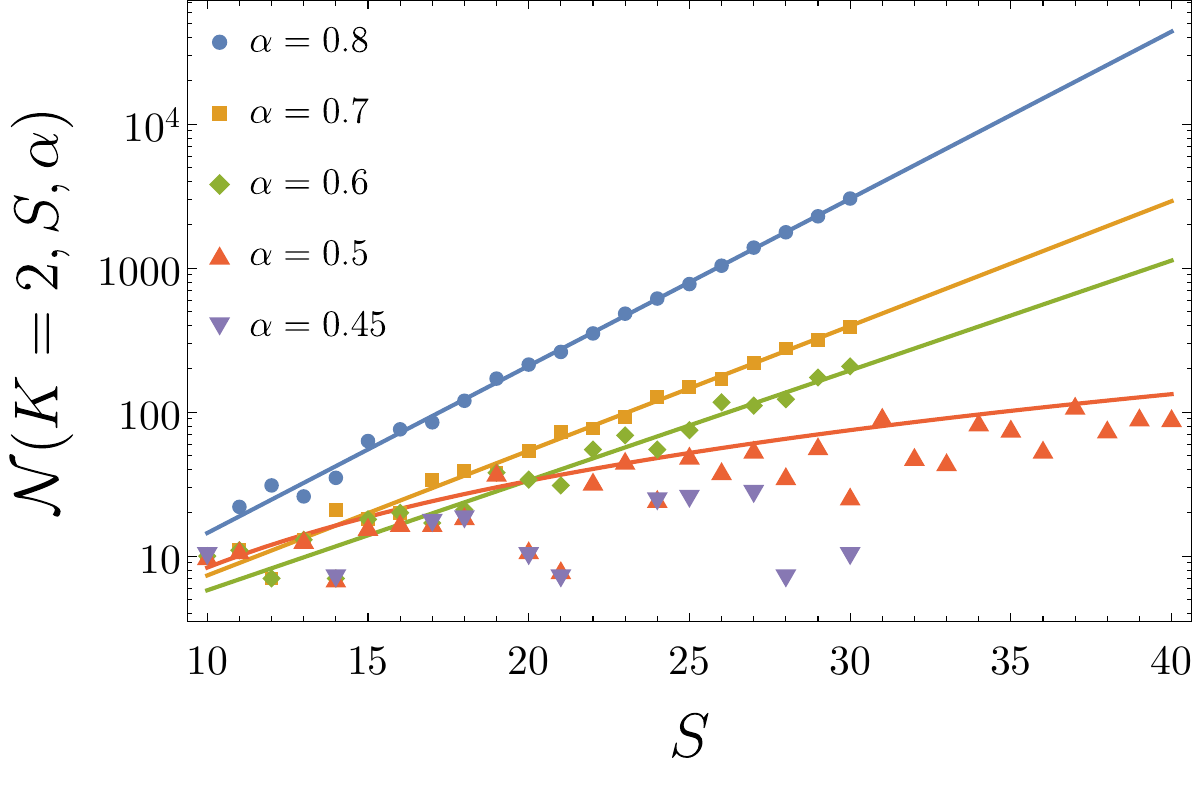}
    \caption{The number of stable fixed points at $K=2$ for different $S$ and $\alpha$. The numbers from exact computations (points) and the asymptotics for $\alpha\geq1/2$ (curves) are shown.}
    \label{fig:count}
\end{figure}

We can continue using this transfer matrix approach to compute the
scaling of the number of solutions as $\alpha$ decreases.  Focusing on
the case $K = 2$, we can determine explicitly which block types are
allowed, and which adjacent pairs of blocks are possible, for various
values of $\alpha$ based on stability and uninvadability:

$\bullet\; 1 >\alpha >1/\sqrt{2}$:
In this region, each block contains a cluster with size $n=1,2,3$, together with
an extinct species on each side. We denote these blocks by
(1), (2), (3) respectively.
All sequences are allowed except those with adjacent (3)(3)'s.
The transfer matrix approach gives
\begin{equation}
 {\cal N} (2, S, 1 >\alpha >1/\sqrt{2}) \sim 1.30638^S\simeq e^{0.267 S}\,.
\label{eq:}
\end{equation}

$\bullet\; 1/\sqrt{ 2}> \alpha > (\sqrt{ 5}-1)/2 \cong 0.618$:
In this region, we can
only have sequences of (1) and (2), with any adjacent pairs allowed, thus
\begin{equation}
 {\cal N} (2, S, 1/\sqrt{ 2}> \alpha >  0.618) \sim
 1.22074^S
 \simeq e^{0.199 S}\,.
\label{eq:}
\end{equation}

$\bullet\; (\sqrt{ 5}-1)/2 \cong 0.618> \alpha > 0.555$:
At $\alpha = (\sqrt{ 5}-1)/2 \cong 0.618$, the sequence (1)(2) becomes 
invadable. At the same time, a chain containing two clusters of size 2
separated by a single extinct species, denoted by (22), becomes stable. This is the first appearance of a chain of interacting clusters as
$\alpha$ decreases. In this
region, we can have sequences of (1), (2), and (22) but not adjacent
(1)(2), thus
\begin{equation}
 {\cal N} (2, S, 0.618> \alpha > 0.555) \sim
 1.19211^S
 \simeq e^{0.176 S}\,.
\label{eq:}
\end{equation}

As $\alpha$ decreases, the scaling continues to go as $\mathcal N \sim
e^{\lambda S}\sim (1+
\epsilon)^S$, with $\epsilon, \lambda \rightarrow 0$ as $\alpha \rightarrow
1/2$.

At $\alpha = 1/2$, as described in Section~\ref{subsec:alpha1/2}, we
have a degenerate 1-parameter family of solutions for any $z = 2 p/q$ with
\begin{equation}
 \frac{S}{2 K}< p < \frac{S}{K +1} \,,
\label{eq:p-bounds}
\end{equation}
and the number of inequivalent translations of this solution
is $S/\mathrm{gcd}(p,S)$. There is also a multiple-parameter family of solution with no extinction if $p=S/2K$ or $p=S/(K+1)$. Thus, the number of solutions goes as
\begin{equation}
 {\cal N} (K, S, \alpha = 1/2) \lesssim \frac{S^2(K-1)}{2K(K+1)} \,,
\label{eq:}
\end{equation}
which saturates when $S$ has few prime factors. 

We can also use numerical methods to explicitly compute the number of
solutions ${\cal N} (K, S, \alpha)$ for various values of the parameters.
For $S \leq 30$, we can identify all solutions by exhaustively testing
each possible subset of surviving species.  At the critical point
$\alpha = 1/2$, we can explicitly construct all solutions for any $S$ from the set
of possible values of $p$ satisfying Eq. (\ref{eq:p-bounds}).  The results for $K=2$ are shown in
Fig. \ref{fig:count}. We see that for $\alpha\geq1/2$, the numbers from
computations match well with the predicted asymptotics, even
at relatively small $S$.
We can also see the general feature that ${\cal  N}$
decreases as $\alpha$ decreases, for any fixed value of $S$.


\bibliography{references}

\end{document}